\documentclass{article}

\usepackage{amssymb,amsfonts,amsmath,stmaryrd}
\usepackage{cite,enumerate,float,indentfirst}
\usepackage{color}

\def\be{\begin{eqnarray}}
\def\ee{\end{eqnarray}}
\def\nn{\nonumber}

\newcommand{\beq}{\begin{equation}}
\newcommand{\eeq}{\end{equation}}
\newcommand{\beqa}{\begin{eqnarray}}
\newcommand{\eeqa}{\end{eqnarray}}
\newcommand{\CR}{\nonumber \\}

\newcommand{\floor}[1]{\lfloor#1\rfloor}
\newcommand{\lam}{\lambda}
\newcommand{\m}{\mu}

\definecolor{red}{rgb}{1,0,0}
\definecolor{orange}{rgb}{1,0.5,0}
\definecolor{violet}{rgb}{0.7,0,1}

\hoffset= - 1.0in         

\begin{document}

\title{
\LARGE{ \bf On a complete solution of the quantum Dell system
}}

\author{
{\bf Hidetoshi Awata$^a$}\footnote{awata@math.nagoya-u.ac.jp},
\ {\bf Hiroaki Kanno$^{a,b}$}\footnote{kanno@math.nagoya-u.ac.jp},
\ {\bf Andrei Mironov$^{c,d,e}$}\footnote{mironov@lpi.ru; mironov@itep.ru},
\ and \  {\bf Alexei Morozov$^{f,d,e}$}\thanks{morozov@itep.ru}
\date{ }
}

\maketitle

\vspace{-5.4cm}

\begin{center}
\hfill FIAN/TD-19/19\\
\hfill IITP/TH-21/19\\
\hfill ITEP/TH-37/19\\
\hfill MIPT/TH-19/19
\end{center}

\vspace{2.7cm}

\begin{center}
$^a$ {\small {\it Graduate School of Mathematics, Nagoya University,
Nagoya, 464-8602, Japan}}\\
$^b$ {\small {\it KMI, Nagoya University,
Nagoya, 464-8602, Japan}}\\
$^c$ {\small {\it Lebedev Physics Institute, Moscow 119991, Russia}}\\
$^d$ {\small {\it ITEP, Moscow 117218, Russia}}\\
$^e$ {\small {\it Institute for Information Transmission Problems, Moscow 127994, Russia}}\\
$^f$ {\small {\it MIPT, Dolgoprudny, 141701, Russia}}
\end{center}

\vspace{.0cm}

\begin{abstract}
The mother functions
for the eigenfunctions of the Koroteev-Shakirov
version of quantum double-elliptic (Dell) Hamiltonians
can be presented as infinite series in Miwa variables,
very similar to the recent conjecture due to J. Shiraishi.
Further studies should clear numerous remaining obstacles
and thus solve the long-standing problem
of explicitly constructing a Dell system,
the top member of the Calogero-Moser-Ruijsenaars system,
with the $PQ$-duality fully explicit at the elliptic level.
\end{abstract}

\section{Introduction}

The systems with both coordinates and momenta lying on two independent tori called {\bf double elliptic systems} (Dell) were introduced in \cite{BMMM}
for a description of the $6d$ Seiberg-Witten theories containing the adjoint matter hypermultiplet. A celebrated property of these systems is self-duality \cite{Ruid,pqdual,BMMM,GM,MM,M}, which, in nowadays terms, is often referred to as the spectral (self)-duality \cite{specdu}.
Relation of these systems to topological strings and extension to the $6d$ Nekrasov functions was later discussed in \cite{MMZ}. Moreover, in \cite{KZALE},  using solutions to the elliptic Knizhnik--Zamolodchikov equations, we discussed the modular properties of these $6d$ gauge theories described by Dell systems and derived in \cite{AMM}.

 One of the problems with the Dell systems is that they are unambiguously defined
 only in the $SU(2)$ case (two particles), while the $SU(n)$ ($n$-particle) generalization admits two formulations described respectively in \cite{MM} and \cite{BH} whose interrelation remains unclear so far.
 Calculations proved to be very tedious:
finding explicit formulas is a non-trivial quest even in the classical case \cite{ABMMZ}, the Nekrasov ($\Omega$-background) case is even more involved
(in Seiberg-Witten theory, these formulas are supposed to describe
the intermediate case of the Nekrasov-Shatashvili limit \cite{NSl,BS}
on the way to the full-fledged Dell deformation of Nekrasov functions \cite{MMZ}).
A new suggestion for the Dell Hamiltonians was done recently in \cite{KS}; it looks close to the older variant in \cite{BH}. Even if not being fully adequate (see reasons below), its simplicity gives a chance to develop further techniques of Dell studies, the main goal of the present paper.

In \cite{KS}, only a simple explicit example of the eigenfunction was considered
in the form of a few lowest terms of expansion in one of the elliptic parameters and only
for the simplest partition $[1]$ (see \cite[eq.(4.19) and Appendix A]{KS}). We are not going to literally extend this result to higher partitions, because an
exact status of the Koroteev--Shakirov (KS) Hamiltonians remains unclear as yet. Instead, we concentrate on a general approach to the eigenfunction problem.  It could seem to be very hard, but luckily this is not the case.
The problem is drastically simplified if one considers \cite{NS} not
Macdonald-like functions {\it per se},
but their much simpler continuations from the Young diagrams (or partitions) $\lambda_i$
to arbitrary values of spectral parameters $y_i$ \cite{DIM}.
Such a function ${\cal M}\{\vec x|\vec y\}$ with the property
\be
{\cal M}\!\left\{y_i=q^{ \lambda_i}t^{n-i}\Big|x_i\right\} = {\rm Mac}_\lambda[x_i],
 \ \ \ \ i=1,\ldots, n
\label{Fou}
\ee
was nicknamed {\it mother function} in \cite{Smir} (see also earlier papers \cite{Ok}),
where an elliptic (rather than generic Dell) version was studied
and related to the theory of elliptic quantum toroidal algebras \cite{ellDIM}.
The main tool in \cite{Smir} were Kostka matrices $K_{\lambda\mu}$,
which describe a triangular transform from generalized Macdonald polynomials to products of the Schur polynomials
thus having two Young diagram indices and a slightly different mother function
depending on a doubled set of $y$-variables,
${\cal K}\{\vec y,\vec y\,'\}$ \cite{Smir}.
Its Dell version, as well as the Kerov deformations
\cite{Ker,MMkerov1} still remain to be built.
It would be interesting to see whether
the two functions, the triangular ${\cal K}$ and the symmetric ${\cal M}$
are related in the Dell case.

The mother function ${\cal M}$ is symmetric in $\{\vec x\}$,
but it is not a polynomial being
rather a formal series in arbitrary negative powers of $x$.
A nice explicit example of such a series is provided by an
elegant generalization of the Macdonald polynomials
introduced recently by J. Shiraishi \cite{S}.
The Shiraishi series are explicitly self-dual, namely, symmetric
under permutation of $\vec x$ and $\vec y$ variables, they have a proper elliptic limit,
being therefore natural candidates for eigenfunctions of
the Dell system. Moreover, the Shiraishi series for the partition $[1]$ and $n=2$
is an eigenfunction of the Dell Hamiltonian (which we verified for few first terms of expansion).
Some problems remain with an explicit realization of the Shiraishi functions as Dell eigenfunctions,
and with probing its various limits.
However, a complete solution of these problems will
complete the program of explicit construction of the Dell systems,
at least in the Nekrasov-Shatashvili limit.

In fact, the very idea of the mother function is quite old, and goes back to the notion of quantum momentum-coordinate ($PQ$-) duality, which implicitly appeared in the S. Ruijsenaars paper \cite{Ruid} and was later discussed in \cite{Etin,GM,M}. While the classical $PQ$-duality is realized just in terms of Hamiltonians and their canonical transformations \cite{M} and is sometimes realized as a gauge transformation within the Hamiltonian reduction \cite{GM}, or, equivalently, in terms of dynamics of zeroes \cite{BB,MM} of the $\tau$-functions of integrable KP/Toda hierarchies \cite{zeroes}, the quantum duality requires the eigenvalue problem, i.e. Hamiltonians must be accompanied by eigenfunctions from the very beginning. That is, if the eigenvalue problem for a Hamiltonian $\hat H_x$, which is an operator acting on the variable $x$, reads
\be
\hat H_x\cdot\Psi_\lambda(x)=E(\lambda)\Psi_\lambda(x)
\ee
then the dual Hamiltonian acts on the variable $\lambda$:
\be
\hat H_\lambda^D\cdot\Psi_\lambda(x)=E^D(x)\Psi_\lambda(x)
\ee
Here $E$ and $E^D$ are some fixed functions of the respective variables $x$ and $\lambda$. These functions are given by the momenta dependence of classical free Hamiltonians  \cite{BMMM,M}:
\be\label{main2}
E(\lambda)=\hat H_x^{free}\Big|_{i\partial_x\to\lambda}
\ee
In the case of many-body integrable system we have several coordinates $x_i$, $i=1,...,n$
and the corresponding $\lambda_i$ are associated with the separated variables provided a system allows such a separation.
Integrability implies that in this case there are $n$ commuting Hamiltonians
and $n$ dual Hamiltonians.
In this context one naturally considers the eigenfunction $\Psi_\lambda(x)$ as a function of the two continuous variables $x$ and $\lambda$. Such a function provides a reference example of the mother function. In the case of the Hamiltonians from the Calogero-Moser-Ruijsenaars-Shneider family, the most informative are the Hamiltonians of the Dell system, which are elliptic both in coordinates and momenta, and are self-dual, i.e. $\hat H_k=\hat H_k^D$; constructing their eigenfunctions is the main goal of this paper.

The main unresolved problem within this approach (seen already at the classical level) is that
it seemingly fail to directly reproduce the most interesting version of the
Dell Hamiltonians in \cite{MM,BGOR,ABMMZ,AMM}.
The claim in those papers was that in order to absorb the symplectic geometry of the problem
and enjoy the explicit $PQ$-duality,
the period matrices of the underlying Seiberg-Witten spectral curves should be dynamical
(momentum) variables rather than just constants.
This problem was mostly ignored (though mentioned) in \cite{KS},
which suggested studying just naive double-periodic Hamiltonians
with no clear relation to the $PQ$-duality,
but our new proposal to build eigenfunctions from (extension of) the {\it self-dual} Shiraishi
functions could restore the relation.
Still, the puzzle of dynamical period matrices persists,
though it is a separate problem to find their quantization (Baxter $Q$-operators),
which is straightforward with the naive choice of \cite{KS}.
The resolution of all these problems can be the original suggestion of \cite{BMMM}
to use the projection method and obtain a dynamical period matrix at genus $n-1$
from a constant period matrix at genus $n$.
The crucial point here should be a peculiarly simple geometry of the
Dell spectral curve, which is a simple aggregation of a few tori.
Another task is a further generalization from the naive Dell system (\ref{Fou}),
related to quantum toroidal algebras,
to arbitrary systems of Bethe-anzatz roots
associated with arbitrary quivers and Nakajima varieties \cite{OkB},
of which the instanton moduli space is a simple example.

\paragraph{Plan of the paper.} Below, we briefly repeat the basics of the Shiraishi-series theory of
mother functions and then discuss their possible role
as eigenfunctions of the Koroteev-Shakirov Hamiltonians.
We consider in more detail the simplest two-particle case $n=2$, while the $n$-particle case in terms of the Shiraishi functions has to be understood as a representation in terms of separated variables of the Dell eigenfunctions.
Then, in two Appendices, we show the relation of the Shiraishi function
to the partition function of supersymmetric gauge theories.
We expect both relations to follow from a relation of Dell
integrable system to six-dimensional supersymmetric gauge theory, and,
in this sense, they provide another, more physical evidence that the (extension of) Shiraishi function solves
the quantum Dell system. These relations also reveal a geometric
interpretation of theoretical meaning of the Shiraishi function.
It is desirable to understand the Shiraishi function from the representation
theory of the Ding-Iohara-Miki (DIM, quantum toroidal) algebra.

\paragraph{Notation.}

We define the odd $\theta$-function
\be
\theta_p(z):={1\over\sqrt{z}}(z;p)_\infty (p/z;p)_\infty (p;p)_\infty={1\over\sqrt{z}}\sum_{k\in\mathbb{Z}}(-1)^kz^{k}p^{k^2/2-k/2}
\label{theta}
\ee
and the even $\theta$-function
\be\label{thp}
\theta^{(e)}_p(z):=\sum_{k\in\mathbb{Z}}z^kp^{k^2}
\ee
with the properties
\be\label{thetap}
\theta_p(z)=-\theta_p(z^{-1}),\ \ \ \ \ \ \
\theta^{(e)}_p(z)=\theta^{(e)}_p(z^{-1}),\ \ \ \ \ \ \ \theta^{(e)}_p(z/w)=z\theta^{(e)}_p(1/(zw))
\ee
Here the Pochhammer symbol is
\be\label{Poch}
(x;q)_p:=\prod_{n=0}^{p-1}(1-q^nx)={(x;q)_\infty\over (q^nx;q)_\infty}
\ee
and
\be
(x;q_1,q_2)_\infty:=\prod_{n,m=0}^{\infty}(1-q_1^nq_2^mx)
\ee
In the standard notation of \cite{BEr,Mum}, $\theta^{(e)}_p(z)=\theta_3(v,\tau)=\theta_{00}(v,\tau)$
with $p=e^{\pi i\tau}$, $z=e^{2\pi iv}$, while changing the $\theta$-function argument $z\to {z\over p}$  (see section 5) makes it $\theta_2=\theta_{10}$.

\section{Mother functions}

To understand the notion of mother function, one should begin from the case of
Schur polynomials.
In $x$-variables, they are extremely simple, for the Young diagram
$R=\{R_1\geq R_2 \geq \ldots\}$
\be
{\rm Schur}_{R}[x_1,\ldots,x_n] =
\frac{\sum_{\sigma \in S_n}  (-)^\sigma \prod_{i=1}^n x_i^{R_\sigma(i)+n-i}}
{\prod_{i<j} (x_i-x_j)}
\label{Schurxy}
\ee
For the one-row Young diagrams, we get just
\be
\frac{x_1^{R_1+1}x_2^{R_2} - x_1^{R_2}x_2^{R_1+1}}{x_1-x_2}
= x_1^{R_1}x_2^{R_2} \cdot
\frac{   1-\left(\frac{x_2}{x_1}\right)^{R_1+1-R_2}\!\!\!\!\!\!\!\!\!\!\!\!}
{\!\!\!\!\!\!\!\!1-\frac{x_2}{x_1}}
\ee
An obvious analytic continuation from integer to arbitrary $R_i$ is provided by just
the same expression, and there is an explicit symmetry $\log x_i$ and
$R_i=\log y_i+i-n$:
after division by an $R$-dependent factor (\ref{Schurxy}) becomes
\be
{\rm Schur}_{R}[x_1,\ldots,x_n] \sim
\frac{\sum_{\sigma \in S_n}  (-)^\sigma \prod_{i=1}^n
e^{\log x_i\cdot \log y_i}}
{\prod_{i<j} (x_i-x_j)(y_i-y_j)}
\ee
However, with this continuation the powers of $x$-variables can be non-integer.
An alternative continuation,
which is assumed in the definition of {\it mother function}
leaves all the powers integer,
but converts a finite polynomial into an infinite series.
The idea is to take
\be
\frac{1-x^{R+1}}{1-x}=\sum_{k=1}^R x^k \ \  \longrightarrow  \ \
\lim_{\epsilon\longrightarrow 0}\left\{\sum_{k=0}^\infty
\frac{\Gamma(k-R)}{\Gamma(-R)}\frac{\Gamma(-R+\epsilon)}{\Gamma(k-R+\epsilon)}\,x^k
\right\}
\ee
The r.h.s. is a hypergeometric function $_1F_1$, which can be bosonised by the method
of \cite{MV}, but we do not need these details here.
What is important, at integer $R$, only the first $R+1$ items in the sum are non-vanishing,
while, at non-integer $R$, one gets an infinite sum with $R$-independent unit coefficients.
Note that the function becomes a symmetric function of $x_i$ only at integer $R_i$ and with the common factor
$\prod_{i=1}^n x_i^{R_i+1}$ inserted.

This singular-looking construction gets automatically regularized already in the
case of Macdonald polynomials,  where $\epsilon$ is no-longer vanishing,
but is rather equal to $\log(q/t)$, thus no limits are needed,
and the coefficients, while still vanishing at appropriate integer $R_i$,
become smooth functions of $R_i$.
This continues to work nicely when the
Macdonald polynomials are further deformed to elliptic Shiraishi series
and their double-elliptic generalizations.

\section{Noumi-Shiraishi representation of Macdonald polynomials}

We continue with the simplest healthy example of the mother function: the case of ordinary Macdonald polynomials. This example was described in detail in \cite{NS}.

Suppose $t^k\notin q^{\mathbb{Z}}$ for $k=1,\ldots, n-1$. For $i,j=1,\ldots,n$ define a power series
\be\label{P}
P_n(x_i,y_i|q,t):=\sum_{m_{ij}}C_n(m_{ij},y_i|q,t)\prod_{1\le i<j\le n}\left({x_j\over x_i}\right)^{m_{ij}}
\ee
where $m_{ij}=0$ for $i\ge j$, $m_{ij}\in \mathbb{Z}_{\ge 0}$,
\be\label{c}
C_n(m_{ij},y_i|q,t):= \nn \\
= \prod_{k=2}^n\prod_{1\le i<j\le k}{\Big(q^{\sum_{a>k}(m_{ia}-m_{ja})}ty_j/y_i;q\Big)_{m_{ik}}
\over \Big(q^{\sum_{a>k}(m_{ia}-m_{ja})}qy_j/y_i;q\Big)_{m_{ik}}}\cdot
\prod_{k=2}^n\prod_{1\le i\le j<k}{\Big(q^{-m_{jk}+\sum_{a>k}(m_{ia}-m_{ja})}qy_j/ty_i;q\Big)_{m_{ik}}
\over \Big(q^{-m_{jk}+\sum_{a>k}(m_{ia}-m_{ja})}y_j/y_i;q\Big)_{m_{ik}}}
\ee
This $P_n(x_i,y_i|q,t)$ solves the eigenvalue problem
\be
\hat D(u) \cdot x^\lambda P_n(x_i,y_i|q,t)=\prod_{i=1}^n (1-uy_i)\cdot x^\lambda P_n(x_i,y_i|q,t)
\ee
where $\lambda$ is a set of complex parameters defined through $q^{\lambda_i}:=y_it^{i-n}$ and
\be
\hat D(u):=\sum_r (-u)^r\hat H_r
\ee
is the generating function of the Ruijsenaars Hamiltonians $\hat H_r$,
\be\label{MHam}
\hat H_r:=t^{n(n-1)/2}\sum_{|I|=r}\prod_{i\in I;j\notin I}{tx_i-x_j\over x_i-x_j}\prod_{i\in I}\hat T_{q,x_i}
\ee
where $\hat T_{q,x_i}f(x_1,\ldots,x_i,\ldots,x_n):=f(x_1,\ldots,qx_i,\ldots,x_n)$.

With the choice $y_i=q^{R_i}t^{n-i}$, the infinite series (\ref{P}) becomes a Laurent,
polynomial proportional to the Macdonald polynomial for the partition $R$ with $l_R=n$,
\be\label{MP}
{\rm Mac}_{_R}(x_i;q,t)=x^R\cdot P_n(x_i,q^{R_i}t^{n-i}|q,t)
\ee

\paragraph{Limit to the Schur polynomials.}
 As already mentioned, the limit of this representation of the Macdonald polynomials to the Schur polynomials
is not naive, since naively the mother function  at $t=q$ does not depend on $y_i$ at all.
The role of the numerator of the first factor in (\ref{c}) is that, when specializing to the Macdonald point $y_i=q^{R_i}t^{n-i}$, it selects out the domain of values of variables $m_{ij}$: the factor
\be
\prod_{k=2}^n\prod_{1\le i<j\le k}\Big(q^{\sum_{a>k}(m_{ia}-m_{ja})}ty_j/y_i;q\Big)_{m_{ik}}
\ee
is non-vanishing iff non-vanishing is the factor with $j=i+1$,
i.e. $0\le m_{ik}\le R_i-R_{i+1}-\sum_{a>k}(m_{ia}-m_{i+1,a})$ for all $1\le i< k$, $1<k\le n$.
However, if one immediately puts $t=q$ in (\ref{c}),
this numerator does not work this way any longer.
Hence, in contrast to the Macdonald case, when one can ascribe
arbitrary complex values to the variables $y_i$,
one can not consider the Schur polynomial outside the values
associated with a concrete Young diagram.
In this case, one has to restrict the admissible values of $m_{ij}$ by hands,
and only after this put $t=q$, what leads to $C_n(m_{ij},y_i|q,t)=1$.
Thus, one obtains in the Schur limit, instead of (\ref{MP}), the expression
\be
{\rm Schur}_{_R}(x_i):=x^R\cdot \sum_{m_{ij}\in{\cal A}_R}
\prod_{1\le i<j\le n}\left({x_j\over x_i}\right)^{m_{ij}}
\ee
where ${\cal A}_R$ is a set of $m_{ik}:\ 0\le m_{ik}\le R_i-R_{i+1}-\sum_{a>k}(m_{ia}-m_{i+1,a})$ for all $1\le i< k$, $1<k\le n$.

\paragraph{Parameterizing $m_{ij}$ by Young diagrams.} The formulas for ${\cal A}_R$ in the previous paragraph suggest to introduce, instead of $m_{ij}$,
\be\label{mu}
\mu_i^{(k)}:=R_i-\sum_{a>k}m_{ik},\ \ \ \ \ i=1,\ldots,k
\ee
Then, the conditions defining ${\cal A}_R$ are nothing that a requirement for $\mu^{(j)}$ to be a set of Young diagrams (with $j$ lines, $j=1,\ldots,n$). In fact, formula (\ref{mu}) follows from the equation
\be\label{mmu}
m_{ij}=\mu_i^{(j)}-\mu_i^{(j-1)}
\ee
and the initial conditions $\mu_i^{(n)}=R_i$, which one additionally imposes when associating $P(x_i,y_i|q,t)$ with the Macdonald polynomial. Generally, it is sufficient to define  $\mu^{(j)}_i$ with (\ref{mmu}).

In particular, with this definition (\ref{mmu}), the $x$-dependent factor in (\ref{P}) can be rewritten in the form
\be
\prod_{1\le i<j\le n}\left({x_j\over x_i}\right)^{m_{ij}}=\prod_{1\le i< n}\left({x_{i+1}\over x_i}\right)^{\sum_{a=1}^i
(\mu_a^{(n)}-\mu_a^{(i)})}
\ee
Choosing  the initial conditions $\mu_i^{(n)}=0$, one arrives at
\be
\prod_{1\le i<j\le n}\left({x_j\over x_i}\right)^{m_{ij}}=\prod_{1\le i< n}\left({x_{i+1}\over x_i}\right)^{-\sum_{a=1}^i
\mu_a^{(i)}}=\prod_{1\le i< n}\prod_{a=1}^i\left({x_{i+1}\over x_i}\right)^{-\mu_a^{(i)}}
\ee
Note that one can define $\Lambda_i^{(j)}:=-\mu_j^{(i)}$ so that $m_{ij}=\Lambda^{(i)}_{j-1}-\Lambda^{(i)}_{j}$, and the condition of non-negativity of $m_{ij}$ would just mean that $\Lambda^{(i)}$ is a Young diagram. However, there is still an additional condition for $\Lambda^{(j)}_i$ that $j\le i$ (see (\ref{mu})). In order to remove it for having an unconstrained set of the Young diagrams, we define, for future convenience, $\lambda^{(i)}_j:=\Lambda^{(i)}_{i+j-1}$ so that the additional condition becomes just $j\ge 1$.
With this definition, the previous factor can be rewritten as
\be
\prod_{1\le i<j\le n}\left({x_j\over x_i}\right)^{m_{ij}}=
\prod_{b=1}^{n-1}\prod_{a=1}^{n-b}\left({x_{a+b}\over x_{a+b-1}}\right)^{\lambda_{a}^{(b)}}
\ee
Note that the initial conditions $\mu_i^{(n)}=0$ reads that the number of lines of the Young diagram $l_{\lambda^{(i)}}\le n-i$. Note also that ${\cal A}_R$ in these variables is:
\be
{\cal A}_R:\ \ \ \ \lambda_j^{(i)}\le R_i-R_{i+1}+\lambda^{(i+1)}_j,\ \ 1\le j\le n-i
\ee

\subsection*{Examples}

\paragraph{\fbox{$n=2$}}
The coefficient (\ref{c}) is
\be
C_2(m_{12},y_1,y_2|q,t)={(ty_2/y_1;q)_{m_{12}}\over (qy_2/y_1;q)_{m_{12}}}
{(t;q)_{m_{12}}\over (q;q)_{m_{12}}}\left({q\over t}\right)^{m_{12}}
\ee
The Macdonald polynomial associated with the 2-line Young diagrams is
\be
{\rm Mac}_{_{[R_1,R_2]}}(x_1,x_2;q,t)=x_1^{R_1}x_2^{R_2}\ \sum_{m=0}^{R_1-R_2}{(q^{R_2-R_1};q)_m\over (q^{R_2-R_1+1}/t;q)_m}
{(t;q)_m\over (q;q)_m}\left({qx_2\over tx_1}\right)^m
\ee
and the corresponding Schur polynomial is
\be
{\rm Schur}_{_{[R_1,R_2]}}(x_1,x_2)=x_1^{R_1}x_2^{R_2}\ \sum_{m=0}^{R_1-R_2}\left({x_2\over x_1}\right)^m
\ee

\paragraph{\fbox{$n=3$}}
The coefficient (\ref{c}) is (notice a misprint in \cite{NS})
\be
C_3(m_{12},m_{13},m_{23},y_1,y_2,y_3|q,t)={(q^{m_{13}-m_{23}}ty_2/y_1;q)_{m_{12}}\over
(q^{m_{13}-m_{23}}qy_2/y_1;q)_{m_{12}}}{(ty_2/y_1;q)_{m_{13}}\over (qy_2/y_1;q)_{m_{13}}}{(ty_3/y_1;q)_{m_{13}}\over (qy_3/y_1;q)_{m_{13}}}
{(ty_3/y_2;q)_{m_{23}}\over (qy_3/y_2;q)_{m_{23}}}\times\nn\\
\times {(q^{-m_{23}}qy_2/ty_1;q)_{m_{13}}\over (q^{-m_{23}}y_2/y_1;q)_{m_{13}}}
{(t;q)_{m_{12}}\over (q;q)_{m_{12}}}{(t;q)_{m_{13}}\over (q;q)_{m_{13}}}{(t;q)_{m_{23}}\over (q;q)_{m_{23}}}
\left({q\over t}\right)^{m_{12}+m_{13}+m_{23}}
\ee
The Macdonald polynomial associated with the 3-line Young diagrams is
\be
{\rm Mac}_{_{[R_1,R_2,R_3]}}(x_i;q,t)=x_1^{R_1}x_2^{R_2}x_3^{R_3}\ \sum_{m_{13}=0}^{R_1-R_2}\ \sum_{m_{23}=0}^{R_2-R_3}\
\sum_{m_{12}=0}^{R_1-R_2+m_{23}-m_{13}}
C_3(m_{12},m_{13},m_{23},t^2q^{R_1},tq^{R_2},q^{R_3}|q,t)\times \nn\\
\times\left({x_2\over x_1}\right)^{m_{12}}\left({x_3\over x_1}\right)^{m_{13}}
\left({x_3\over x_2}\right)^{m_{23}}
\ee
and the corresponding Schur polynomial is
\be
{\rm Schur}_{_{[R_1,R_2,R_3]}}(x_i)=x_1^{R_1}x_2^{R_2}x_3^{R_3}\ \sum_{m_{13}=0}^{R_1-R_2}\ \sum_{m_{23}=0}^{R_2-R_3}\
\sum_{m_{12}=0}^{R_1-R_2+m_{23}-m_{13}}
\left({x_2\over x_1}\right)^{m_{12}}\left({x_3\over x_1}\right)^{m_{13}}
\left({x_3\over x_2}\right)^{m_{23}}
\ee
This same expression in the $\lambda^{(i)}$-variables is
\be
{\rm Schur}_{_{[R_1,R_2,R_3]}}(x_i)=x_1^{R_1}x_2^{R_2}x_3^{R_3}\ \sum_{\lambda_1^{(2)}=0}^{R_2-R_3}\ \sum_{\lambda_2^{(1)}=0}^{R_1-R_2}\
\sum_{\lambda_1^{(1)}=\lambda_2^{(1)}}^{R_1-R_2+\lambda_1^{(2)}}
\left({x_2\over x_1}\right)^{\lambda_1^{(1)}}\left({x_3\over x_2}\right)^{\lambda_1^{(2)}+\lambda_2^{(1)}}
\ee

\section{Shiraishi functions \label{S}}

Now we are ready to describe the double deformation of the Noumi-Shiraishi representation of the Macdonald polynomials, which was proposed by J. Shiraishi \cite{S}, who constructed it as an average of the product of screened vertex operators made of the affine screening operators.

Define
\be\label{ellP}
\mathfrak{P}_n(x_i;p|y_i;s|q,t):=\sum_{\lambda^{(i)}}\prod_{i,j=1}^n{\mathrm{N}^{(j-i)}_{\lambda^{(i)}\lambda^{(j)}}(ty_j/y_i|q,s)
\over \mathrm{N}^{(j-i)}_{\lambda^{(i)}\lambda^{(j)}}(y_j/y_i|q,s)}\cdot
\prod_{b=1}^n\prod_{a\ge 1}\left({px_{a+b}\over
tx_{a+b-1}}\right)^{\lambda_a^{(b)}}
\ee
where $\{\lambda^{(i)}\}$, $i=1,\ldots,n$ is a set of $n$ partitions, we assume that $x_{i+n}=x_i$, and
\be\label{ellC}
\mathrm{N}^{(k)}_{\lambda\mu}(u|q,s):=\prod_{{\beta\ge \alpha\ge 1}\atop{\beta-\alpha=k\ mod\ n}}
\Big(uq^{-\mu_\alpha+\lambda_{\beta+1}}s^{\beta-\alpha};q\Big)_{\lambda_\beta-\lambda_{\beta+1}}
\prod_{{\beta\ge \alpha\ge 1}\atop{\beta-\alpha=-k-1\ mod\ n}}\Big(uq^{\lambda_\alpha-\mu_\beta}s^{\alpha-\beta-1};q\Big)_{\mu_\beta-\mu_{\beta+1}}
\ee
This is what has to do with an eigenfunction of the quantum Dell Hamiltonian \cite{KS}. $\mathfrak{P}(x_i;p|y_i;s|q,t)$ is a symmetric function w.r.t. simultaneous permutations of the pairs $(x_i,y_i)$, however, it is not a symmetric function of $x_i$ only. In order to give rise to a symmetric function of $x_i$, one has to choose this time $y_i=q^{R_i}(ts)^{n-i}$. Then, the function
\be\label{main}
\mathfrak{M}_R(x_i|p,s|q,t):=x^R\cdot \mathfrak{P}_n(p^{n-i}x_i;p|q^{R_i}(ts)^{n-i};s|q,{q\over t})
\ee
is a symmetric function.

\paragraph{Dualities.} J. Shiraishi has conjectured \cite{S} two duality formulas generalizing the corresponding duality formulas for the ordinary Macdonald polynomials: $pq$-duality
\be\label{pqd}
{\mathfrak{P}_n(x_i;p|y_i;s|q,t)\over \mathfrak{P}_n(x_i;p|y_i;0|q,t)}={\mathfrak{P}_n(y_i;s|x_i;p|q,t)
\over \mathfrak{P}_n(y_i;s|x_i;0|q,t)}
\ee
and Poincare duality
\be
{\mathfrak{P}_n(x_i;p|y_i;s|q,t)\over \mathfrak{P}_n(x_i;p|y_i;0|q,t)}={\mathfrak{P}_n(x_i;p|y_i;s|q,{q\over t})
\over \mathfrak{P}_n(x_i;p|y_i;0|q,{q\over t})}
\ee
Note that
\be\label{33}
\mathfrak{P}_n(x_i;p|y_i;0|q,t)=\prod_{1\le i<j\le n}{(p^{j-i}qx_j/x_i;q,p^n)_\infty\over (p^{j-i}tx_j/x_i;q,p^n)_\infty}
\prod_{1\le i\le j\le n}{(p^{n-j+i}qx_j/x_i;q,p^n)_\infty\over (p^{n-j+i}tx_j/x_i;q,p^n)_\infty}
\ee

\paragraph{The limit to the elliptic Ruijsenaars system.}
Another important conjecture by J. Shiraishi deals with the limit to the elliptic Ruijsenaars system. That is,
let $\xi(p|y_i;s|q,t)$ be the constant term of $\mathfrak{P}_n(x_i;p|y_i;s|q,t)$ w.r.t. $x_i$:
\be
\xi(p|y_i;s|q,t):=\sum_{{\lambda^{(i)}}\atop{m_1=\ldots=m_n=0}}\prod_{i,j=1}^n{\mathrm{N}^{(j-i)}_{\lambda^{(i)}\lambda^{(j)}}(ty_j/y_i|q,s)
\over \mathrm{N}^{(j-i)}_{\lambda^{(i)}\lambda^{(j)}}(y_j/y_i|q,s)}\cdot\Big({p\over t}\Big)^{|{\bf\lambda}|}
\ee
where we have introduced the notation: $|{\bf\lambda}|:=\sum_b|\lambda^{(b)}|$, $m_i:=\displaystyle{\sum_b\sum_{{a\ge 1}\atop{a+b=i\ mod\ n}}(\lambda_a^{(b)}-\lambda_a^{(b+1)})}$ (i.e. $|{\bf\lambda}|=0\ mod\ n$). Then, the elliptic counterpart of the Macdonald polynomial is the function (the naive limit of (\ref{ellP}) at $s=1$ is singular)
\be\label{Rui1}
{\cal P}_n(x_i;p|y_i|q,t):=\xi(p|y_i;s|q,t)^{-1}\cdot\mathfrak{P}_n(x_i;p|y_i;s|q,t)\Big|_{s=1}
\ee
It is conjectured \cite{S} to be the eigenfunction of the elliptic Ruijsenaars Hamiltonian:
\be\label{Rui2}
\hat {\cal D}_1\cdot x^\lambda{\cal P}_n(p^{n-i}x_i;p|y_i|q,{q\over t})=\Lambda(y_i|p|q,t)
\cdot x^\lambda{\cal P}_n(p^{n-i}x_i;p|y_i|q,{q\over t}),\nn\\
\hat {\cal D}_1:=t^{n/2}\sum_{i=1}^n\prod_{j\ne i}{\theta_{p^n}(tx_i/x_j)\over\theta_{p^n}(x_i/x_j)}\hat T_{q,x_i}
\ee
Here again, $\lambda$ is a set of complex parameters defined through $q^{\lambda_i}:=y_it^{i-n}$.
Note that $\Lambda(y_i|p|q,t)$ is a power series in $p$, $\Lambda(y_i|0|q,t)=\sum_{i=0}^ny_i$.

\section{Quantum Dell Hamiltonians \cite{KS}\label{KS}}

A quantum counterpart of the Dell Hamiltonians proposed in \cite{KS} is
\be\label{Ham}
\hat {\mathfrak{H}}_a(w,u|q,t):=\hat{\mathfrak{O}}_0^{-1}(w,u|q,t)\cdot\hat{\mathfrak{O}}_a(w,u|q,t),\ \ \ \ \ \ \ a=1\ldots n-1
\ee
where $\hat{\mathfrak{O}}_a$ is read from
\be\label{Oop}
\hat{\mathfrak{O}}(z|w,u|q,t)=\sum_{k\in\mathbb{Z}}\hat{\mathfrak{O}}_k(w,u|q,t)z^k:=\sum_{k_1,\ldots,k_n\in\mathbb{Z}}
z^{\sum_i k_i}w^{\sum_i k_i(k_i-1)/2}\prod_{i<j}\theta_{u^2}\Big(t^{k_i-k_j}x_i/x_j\Big)
\prod_{i=1}^n\hat T_{q,x_i}^{k_i}
\ee
The Hamiltonians $\hat {\mathfrak{H}}_a$ are conjectured to commute with each other (it was checked in \cite{KS} for the first terms with the computer checks). The Hamiltonians depend on two parameters $w$ and $u$ that are associated with the double elliptic deformation.
There is also a trivial Hamiltonian at $a=n$:
\be
\hat {\mathfrak{H}}_n(w,p|q,t)=\prod_{i=1}^n\hat T_{q,x_i}^{k_i}
\ee
Shiraishi functions are trivially its eigenfunctions, since they are graded symmetric functions.

We {\bf conjecture} that (an extension of) the Shiraishi master function (\ref{ellP})-(\ref{ellC}) solves the eigenvalue problem of the Dell Hamiltonians (\ref{Ham})-(\ref{Oop}):
\be\label{main1}
\hat {\mathfrak{H}}_a(w,u|q,t)\cdot x^\lambda\mathfrak{P}_n\!\left(p^{n-i}x_i;p\,\Big|\,y_i;s\,\Big|\,q,{q\over t}\right)=\Lambda_a(y_i|p,s|q,t)
\cdot x^\lambda\mathfrak{P}_n\!\left(p^{n-i}x_i;p\,\Big|\,y_i;s\,\Big|\,q,{q\over t}\right)
\ee
with some identification of parameters $(w,u)\to (s,p)$.
In particular, the limit $s\to 1$ corresponds to $w\to 0$. Note that $\hat {\mathfrak{H}}_a(w,u|1,1)$ become functions at the $q=t=1$ point, and these eigenvalues are dictated by the general rules of \cite{M}. In fact, in the next section, we consider the case of $n=2$ and demonstrate that the Shiraishi function provides an eigenfunction of the Dell Hamiltonian in the case of simplest partition $[1]$, while higher partitions may require an extension.

Now we briefly consider various limits of this formula.

\paragraph{$w\to 0$ limit.} In this limit, the Hamiltonians (\ref{Ham}) reduce to the elliptic Ruijsenaars Hamiltonians, in particular, the first one is $\hat {\cal D}_1$ in (\ref{Rui2}), and, in accordance with (\ref{Rui2}), the Shiraishi function is an eigenfunction of this Hamiltonian provided the $w\to 0$ is associated with $s\to 1$. A typical exact formula is (\ref{ws}).

\paragraph{$u\to 0$ limit.} This limit is dual to the $w\to 0$ limit. Hence, one has to expect that it should correspond to the Shiraishi functions in the $p\to 1$ limit. On the other hand, the Dell Hamiltonians are reduced in this case to the Hamiltonians of the system dual to the elliptic Ruijsenaars one. Its eigenvalues can be explicitly constructed as we discuss in the next section in the two-particle case, the extension to arbitrary number of particles being immediate. As for the $p\to 1$ limit of the Shiraishi function, there are some problems with it.

\paragraph{$p\to 1$ limit.} Indeed, the Shiraishi function is defined as a formal power series in $p$. One may think that it is possible to use the duality (\ref{pqd}) in order to deal with this limit. However,
as follows from (\ref{pqd}), the limit of $\mathfrak{P}_2(p^{n-i}x_i;p|y_1,y_2;s|q,{q\over t})$ at $p\to 1$ is given by
\be
\mathfrak{P}_n(p^{n-i}x_i;p|y_i;s|q,{q\over t})\Big|_{p\to 1}=
\left({\mathfrak{P}_n(p^{n-i}x_i;p|y_i;0|q,{q\over t})\over\mathfrak{P}_n(y_i;s|p^{n-i}x_i;0|q,{q\over t})}\cdot
\mathfrak{P}_n(y_i;s|p^{n-i}x_i;p|q,{q\over t})\right)\Biggr|_{p\to 1}
\ee
Here the $x$-dependent factor $\mathfrak{P}_n(p^{n-i}x_i;p|y_i;0|q,{q\over t})$ is given by (\ref{33}) (note that, in accordance with (\ref{33}), $\mathfrak{P}_n(y_i;s|p^{n-i}x_i;0|q,{q\over t})$ does not depend on $x_i$ and can be removed by changing the normalization) and requires a regularization in the $p\to 1$ limit:
\be
\mathfrak{P}_n(p^{n-i}x_i;p|y_i;0|q,{q\over t})=\prod_{1\le i<j\le n}{(qx_j/x_i;q,p^n)_\infty\over (qx_j/tx_i;q,p^n)_\infty}
\prod_{1\le i\le j\le n}{(p^{n-2j+2i}qx_j/x_i;q,p^n)_\infty\over (p^{n-2j+2i}qx_j/tx_i;q,p^n)_\infty}
\ee
is divergent at $p\to 1$. Hence, dealing with the Shiraishi function in the limit $p\to 1$ is not immediate.

\paragraph{$p\to 0$ limit.}
The limit $p\to 0$ in the Shiraishi functions is the limit to the ordinary Macdonald functions. We discuss it in detail in Appendix A. Note that, in this limit, the $\theta$-functions in (\ref{Ham})-(\ref{Oop}) become just
\be
\theta_p(x)\Big|_{p\to 0}={1-x\over\sqrt{x}}
\ee
The Dell Hamiltonians (\ref{Oop}) have to reduce in this case to the ordinary Macdonald Hamiltonians, i.e. $p\to 0$ limit corresponds to
both $w\to 0$ and $u\to 0$.

In the next section, we discuss our conjecture very explicitly in the case of $n=2$.

\section{Two particle $n=2$ case}

\subsection{Shiraishi functions}

Consider the simplest case of $n=2$. In this case,
\be\label{dellP}
\mathfrak{P}_2(x_1,x_2;p|y_1,y_2;s|q,t)=\sum_{\lambda^{(1)},\lambda^{(2)}}{\mathrm{N}^{(0)}_{\lambda^{(1)}\lambda^{(1)}}(t|q,s)
\over \mathrm{N}^{(0)}_{\lambda^{(1)}\lambda^{(1)}}(1|q,s)}\cdot
{\mathrm{N}^{(0)}_{\lambda^{(2)}\lambda^{(2)}}(t|q,s)
\over \mathrm{N}^{(0)}_{\lambda^{(2)}\lambda^{(2)}}(1|q,s)}
{\mathrm{N}^{(1)}_{\lambda^{(1)}\lambda^{(2)}}(ty_2/y_1|q,s)
\over \mathrm{N}^{(1)}_{\lambda^{(1)}\lambda^{(2)}}(y_2/y_1|q,s)}
{\mathrm{N}^{(1)}_{\lambda^{(2)}\lambda^{(1)}}(ty_1/y_2|q,s)
\over \mathrm{N}^{(1)}_{\lambda^{(2)}\lambda^{(1)}}(y_1/y_2|q,s)}\times\nn
\ee
\vspace{-0.3cm}
\be
\times
\left({p\over t}\right)^{|\lambda^{(1)}|+|\lambda^{(2)}|}
\left({x_{2}\over x_{1}}\right)^{\sum_{i\ge 1}(\lambda_{2i-1}^{(1)}-\lambda_{2i}^{(1)}+\lambda_{2i}^{(2)}-\lambda_{2i-1}^{(2)})}
\ee
Here
\be
\mathrm{N}^{(0)}_{\lambda\lambda}(u|q,s):=\prod_{{j\ge i\ge 1}\atop{j-i=even}}
\Big(uq^{-\lambda_i+\lambda_{j+1}}s^{j-i};q\Big)_{\lambda_j-\lambda_{j+1}}
\prod_{{j\ge i\ge 1}\atop{j-i=odd}}\Big(uq^{\lambda_i-\lambda_j}s^{i-j-1};q\Big)_{\lambda_j-\lambda_{j+1}}\nn\\
\mathrm{N}^{(1)}_{\lambda\mu}(u|q,s):=\prod_{{j\ge i\ge 1}\atop{j-i=odd}}
\Big(uq^{-\mu_i+\lambda_{j+1}}s^{j-i};q\Big)_{\lambda_j-\lambda_{j+1}}
\prod_{{j\ge i\ge 1}\atop{j-i=even}}\Big(uq^{\lambda_i-\mu_j}s^{i-j-1};q\Big)_{\mu_j-\mu_{j+1}}
\ee
Note that potentially there could be factors that vanish at some values of $\lambda^{(1,2)}$: $\mathrm{N}^{(0)}_{\lambda\lambda}(1|q,s)$. However, the both factors that could restrict the admissible values of $\lambda^{(1,2)}$, i.e. when the degree of $s$ is zero, have the form $(q^{-n};q)_n$, which is non-vanishing.
The sum (\ref{dellP}) giving the Dell polynomial is a power series in $p$, which one can manifestly construct term by term. For instance, the constant term is just 1, and the linear term gets contributions when only one of $\lambda^{(1)}_1$ and $\lambda^{(2)}_1$ is non-vanishing and equal to 1. The first terms in $p$ in this expression are
\be
\mathfrak{P}_2(x_1,x_2;p\,|\,y_1,y_2;s\,|\,q,t)
=1+p\cdot{1-qt^{-1}\over 1-q}\left({qsy_1-ty_2\over qsy_1-y_2}{x_1\over x_2}+
{qsy_2-ty_1\over qsy_2-y_1}{x_2\over x_1}\right)+\nn
\ee
\vspace{-0.5cm}
\be
\!\!\!\!\!\!
+\,p^2\cdot\left({1-q^2t^{-1}\over 1-q^2}{1-qt^{-1}\over 1-q}
{q^2sy_1-ty_2\over q^2sy_1-y_2}{qsy_1-ty_2\over qsy_1-y_2}{x_1^2\over x_2^2}+
{1-q^2t^{-1}\over 1-q^2}{1-qt^{-1}\over 1-q}
{q^2sy_2-ty_1\over q^2sy_2-y_1}{qsy_2-ty_1\over qsy_2-y_1}{x_2^2\over x_1^2}+const
\right)+O(p^3)
\nn
\ee
Now, in accordance with (\ref{main}), in order to make a reduction to the symmetric function corresponding to the Young diagram one has, first of all, to make the substitution $x_1\to px_1$, $t\to {q\over t}$:
\be\label{12}
\mathfrak{P}_2\left(p\cdot x_1,x_2;p\,\Big|\,y_1,y_2;s\,\Big|\,q,{q\over t}\right)
= 1+{q\over t}{1-t\over 1-q}{sty_2-y_1\over qsy_2-y_1}{x_2\over x_1}
+{q^2\over t^2}{1-qt\over 1-q^{2}}{1-t\over 1-q}{qsty_2-y_1\over q^2sy_2-y_1}{sty_2-y_1\over qsy_2-y_1}{x_2^2\over x_1^2}
+ \ldots +O(p^2) =
\nn
\ee
\vspace{-0.5cm}
\be
=1+\sum_{k=0}{x_2^{k+1}\over x_1^{k+1}}
\prod_{i=0}^k {q^{i+1}\over t^{i+1}}{1-q^it\over 1-q^{i+1}}{q^isty_2-y_1\over q^{i+1}sy_2-y_1}+p\cdot 0+O(p^2)
\ee
Now one can restrict this function to particular Young diagram $R$.
For instance, for $R=[1]=[1,0]$ we put $y_1=qts$, $y_2=1$
and the series in (\ref{12}) is truncated so that only the first two terms survive, and one obtains
\be
\mathfrak{M}_{[1]}^{(0)}(x_1,x_2|p,s|q,t)=
x_1\cdot\mathfrak{P}_2\left(p\cdot x_1,x_2;p\,\Big|\,qts,1,s\,\Big|\,q,{q\over t}\right)
= x_1+x_2+O(p^2)
\ee
Similarly, in the case of $R=[2]=[2,0]$, one puts $y_1=q^2ts$, $y_2=1$, the series (\ref{12}) is truncated with only the three first terms remaining, and one obtains
\be
\mathfrak{M}_{[2]}^{(0)}(x_1,x_2|p,s|q,t)=x_1^2\cdot\mathfrak{P}_2\left(p\cdot x_1,x_2;p\,\Big|\,q^2ts,1;s\,\Big|\,q,{q\over t}\right)
=x_1^2+x_2^2+{(1-t)(1+q)\over 1-qt}x_1x_2+O(p^2)
\ee
In the case of $R=[1,1]$, one puts $y_1=qts$, $y_2=q$, only the first term in (\ref{12}) remains, and one obtains
\be
\mathfrak{M}_{[1,1]}^{(0)}(x_1,x_2|p,s|q,t)=x_1x_2\cdot\mathfrak{P}_2\left((p\cdot x_1,x_2;p\,\Big|\,qts,q;s\,\Big|\,q,{q\over t}\right)=x_1x_2+O(p^2)
\ee
and so on.

\paragraph{\fbox{$R=[1,0]$}}
Let us consider the simplest Young diagram $R=[1]=\Box$ in more detail.
We collect more terms, the answer looks like
\be\label{M}
\mathfrak{M}_{\Box}^{(0)}(x_1,x_2|p,s|q,t)=x_1\cdot\mathfrak{P}_2\left(p\cdot x_1,x_2;p\,\Big|\,qts,1;s\,\Big|\,q,{q\over t}\right)=\nn
\ee
\vspace{-0.7cm}
\be
=(x_1+x_2)\left[1+
p^2{1-t\over 1-q^2s^2t}\cdot\Big({q\over t}{1-qs^2t^2\over 1-q}{(x_1+x_2)^2\over x_1x_2}+{s^2\over t}
{q-t\over 1-s^2}(2qt+q+t+2)\Big)\right]+O(p^4)
\ee
Note that one can also expand around the point $s=1$.
The function $\mathfrak{P}_2\left(p\cdot x_1,x_2;p,\Big|\,qts,1,s\,\Big|\,q,{q\over t}\right)$
is singular at this point, and we explained in sec.\ref{S} how to choose the proper normalization factor in order to have a smooth limit: one has just to extract the constant term in the brackets in (\ref{M}). Then, after rescaling, the answer has the form
\vspace{-0.3cm}
\be\label{Mt}
{\mathfrak{M}}_{\Box}(x_1,x_2|p,s|q,t)=x_1\cdot\xi\left(p\,\Big|\,qts,1;s\,\Big|\,q,{q\over t}\right)^{-1}\cdot\mathfrak{P}\left(p\cdot x_1,x_2;p\,\Big|\,qts,1;s\,\Big|\,q,{q\over t}\right)
=\nn
\ee
\vspace{-0.6cm}
\be
=(x_1+x_2)\left[1+
p^2{1-t\over 1-q}{q\over t}{1-qs^2t^2\over 1-q^2s^2t}{(x_1^2-x_1x_2+x_2^2)\over x_1x_2}\right]+O(p^4)
\ee
which, indeed, has a smooth limit at $s=1$. Note that it can be written in the form
\be
\!\!
{\mathfrak{M}}_{\Box}(x_1,x_2|p,s|q,t)=\mathfrak{p}_0+
\eta_{001}\Big(p^2{q\over t}\Big)\cdot\left[\eta_{142}\eta_{122}+
p^2\Big(\eta_{142}\eta_{122}\eta_{222}\eta_{121}^{-1}\eta_{221}^{-1}-
{1-q/t\over 1-q^2}{1-q^2s^2t^2\over 1-q^2s^2t}{q-s^2t^2\over 1-qs^2t}\Big)\right]\cdot \mathfrak{p}_1+
\nn
\ee
\vspace{-0.6cm}
\be
+\eta_{001}\Big(p^2{q\over t}\Big)^2\eta_{101}\eta_{122}\eta_{222}
\cdot \mathfrak{p}_2+O(p^6)
\ee
where the time variables are defined as $\mathfrak{p}_k:={\sum_ix_i^{2k+1}\over\prod_i x_i^k}$
\ and \
$\eta_{ijk}:={1-q^is^jt^k\over 1-q^{i+1}s^jt^{k-1}}$.

\subsection{Shiraishi function for the one-box Young diagram as an eigenfunction}

Consider the Dell Hamiltonian in the two-particle, $n=2$ case. In this case,
\be\label{n2hams}
\hat{\mathfrak{O}}_0=\sum_{k\in\mathbb{Z}}
w^{k^2}\theta_{u^2}\Big(t^{2k} {x_1\over x_2}\Big)\hat T_{q,x_1}^{k}\hat T_{q,x_2}^{-k},\ \ \ \ \
\hat{\mathfrak{O}}_1=\sum_{k\in\mathbb{Z}}
w^{k^2-k}\theta_{u^2}\Big(t^{2k-1} {x_1\over x_2}\Big)\hat T_{q,x_1}^{k}\hat T_{q,x_2}^{-k+1}
\ee
and one has to check that ${\mathfrak{M}}_{R}(x_1,x_2|p,s|q,t)$ solves the equation
\be\label{Ham2}
\hat{\mathfrak{O}}_1(u,w|q,t){\mathfrak{M}}_{R}(x_1,x_2|p,w|q,t)-\Lambda_R(p,w|q,t)\cdot
\hat{\mathfrak{O}}_0(u,w|q,t){\mathfrak{M}}_{R}(x_1,x_2|p,w|q,t)=0
\ee
with some eigenvalue $\Lambda_R(p,w|q,t)$. For the one-box Young diagram, it looks so that one can put $u=p$ so that the $p\to 0$ limit is equivalent to the $p\to 1$ limit (which could be the case if there exists a kind of modular invariance relating $p-0$ and $p=1$ points).
We checked this with the computer,
here we list just a few first terms of the $(w,p)$-expansion:
\be
\Lambda_{\Box}(p,w|q,t)=-{qt+1\over t^{1/2}} +w\cdot{(qt+1)(q^2t^2+1)\over t^{3/2}q}-w^2{(qt+1)(q^4t^4+q^3t^3+q^2t^2+qt+1)\over q^2t^{5/2}}+
\nn
\ee
\vspace{-0.6cm}
\be
+w^3{(qt+1)(q^2t^2+1)(q^4t^4+q^3t^3+q^2t^2+qt+1)\over q^3t^{7/2}}+O(w^4)
-p^2\cdot\left({(t-1)(qt+1)(q-t)^2\over t^{3/2}(qt^2-1)}-\right.
\ee
\vspace{-0.6cm}
\be
 \left.
-w\cdot{(t-1)(qt+1)(q-t)^2(t^4q^4-t^3q^4+q^3t^3+q^2t^3+2q^3t+2q^2t^2+2t^3q+q^2t+qt-t+1)\over t^{7/2}q^2(q^2t-1)}+O(w^2)\right)
\nn
\ee
The parameters $s$ and $w$ are related in non-trivial way:
\vspace{-0.1cm}
\be
s-1=2(qt^2-1)(q^2t^2-1)(q^2t-1)\sum_{k=1}\Big({w\over 4q^3t^3}\Big)^k\cdot{\phi_k(q,t)}
\ee

\vspace{-0.4cm}
\noindent with
\vspace{-0.4cm}

{\footnotesize
\be
\phi_1(q,t)=qt+1
\nn
\ee
\vspace{-0.5cm}
\be
\phi_2(q,t)=3q^7t^7+6q^6t^6+q^6t^5+q^5t^6-2q^5t^4-2q^4t^5-3q^4t^4+4q^4t^3+4q^3t^4-q^3t^3+2q^3t^2+2q^2t^3+3q^2t+3qt^2+2qt+1
\nn
\ee
\vspace{-0.5cm}
\be
\phi_3(q,t)=2(qt+1)^2(5q^{12}t^{12}+\ldots)
\nn
\ee
\vspace{-0.5cm}
\be
\ldots\nn
\ee
}
Note that the transformation gets a little bit simpler for the combination $s^2-1$:
\be
s^2-1={(qt^2-1)(q^2t^2-1)(q^2t-1)\over qt}\sum_{k=1}\Big({w\over q^2t^2}\Big)^k\cdot\Phi_k(q,t)
\nn
\ee
Then
\vspace{-0.5cm}
{\footnotesize
\be \ \ \ \ \ \
\Phi_1(q,t)=qt+1
\nn
\ee
\vspace{-0.5cm}
\be
\Phi_2(q,t)=q^6t^6+2q^5t^5-q^4t^3-q^3t^4-q^3t^3+q^3t^2+q^2t^3+q^2t+qt^2+q+t
\nn
\ee
\vspace{-0.5cm}
\be
\Phi_3(q,t)=(qt+1)(q^{10}t^{10}+\ldots)
\nn
\ee
\vspace{-0.5cm}
\be
\ldots \nn
\ee
}

\vspace{-0.5cm}
\noindent
It looks like this relation between $s$ and $w$ does not depend on $p$ and, hopefully, on the Young diagram: hence, it is sufficient to calculate it in the first non-vanishing order in $p$ for the simplest one-box Young diagram.
Thus, one can just substitute into (\ref{Ham2}) the first terms of expansion
\be
\theta_{p^2}(z)={1-z-p^2z^{-1}+z^2p^2\over\sqrt{z}}+O(p^4)
\ee
and use only the terms written down in (\ref{Mt}). In this way, one obtains a series of relations that are satisfied with using the properties (\ref{thp}) of the $\theta$-function and
\be\label{Lambda0}
\Lambda_{\Box}(p,w|q,t)={1\over  \sqrt{t}}{\theta^{(e)}_w(qt/w)\over\theta^{(e)}_w(qt)}+O(p^2)
\ee
from the $p^0$-terms and then
\be\label{Lambda}
\Lambda_{\Box}(p,w|q,t)={1\over  \sqrt{t}}{\theta^{(e)}_w(qt/w)\over \theta^{(e)}_w(qt)}
+p^2\cdot {1\over \sqrt{t}(q-1)\Big(\theta^{(e)}_w(qt)\Big)^2}\cdot
\left[\Xi_{q,t}^{(-1)}+
{\Xi_{q,t}^{(1)}\over
\Xi_{t,q}^{(1)}}\Xi_{t,q}^{(-1)}
\right]+O(p^4)
\ee
\be\label{ws}
s^2={1\over qt}{\Xi_{q,t}^{(1)}-\Xi_{t,q}^{(1)}\over q\Xi_{q,t}^{(1)}-t\Xi_{t,q}^{(1)}}
\ee
\be
\Xi_{q,t}^{(a)}(w):=(q-1)\left(t\theta^{(e)}_w(qt/w)\theta^{(e)}_w(t^3q^a)-q^{(1-a)/2}
\theta^{(e)}_w(qt)\theta^{(e)}_w(t^3q^a/w)\right)
\ee
from the $p^2$-terms.

Thus, the relations that guarantee that (\ref{Ham2}) is correct for the one-box Young diagram and in the first two non-vanishing orders in $p$ fix not only the first terms of $p$-expansion of the eigenvalues (\ref{Lambda}) but also the exact relation (\ref{ws}) between $w$ and $s$.

\subsection{Dual to the elliptic Ruijsenaars system, $u\to 0$}

Formula (\ref{Lambda0}) can be easily generalized to an arbitrary partition:
\be
\Lambda_{[r_1,r_2]}(p,w|q,t)\Big|_{p=0}={q^{r_2}\over  \sqrt{t}}{\theta^{(e)}_w(q^{r_1-r_2}t/w)\over\theta^{(e)}_w(q^{r_1-r_2}t)}
={y_2\over  \sqrt{t}}{\theta^{(e)}_w(y_1/(wy_2))\over\theta^{(e)}_w(y_1/y_2)}
\ee
where $y_1=q^{r_1}t$, $y_2=q^{r_2}$. This formula is consistent with (\ref{main2}), since, in accordance with the general rule (\ref{main2}), one has just to substitute in (\ref{Oop}) $T_{q,x_i}$ for $y_i$ and remove the $x$-dependent factor.

Note that, say, in the case of the first non-trivial partition $R=[2]$, the eigenvalue is given by this formula, but the eigenfunction should be slightly corrected, i.e. the $p\to 0$ limit does not work in this case, and one has probably to consider $p\to 1$:
\be
\!\!\!\!
\Psi_{[n]}(x_1,x_2|p,s|q,t)=M_{[n]}(x_1,x_2|q,t)
-wx_1x_2\cdot
{q\over t^2}{(y_1-ty_2)(y_1-qty_2)\over y_1-qy_2}{(t-1)(q-t)\over  q-1}{y_1+y_2\over y_1}M_{[n-2]}(x_1,x_2|q,qt)+
\nn
\ee
\vspace{-0.5cm}
\be
+O(w^2)+O(p^2)
\label{Macreexpan}
\ee
These expressions should be compared with
\be
\Psi_{[1]}(x_1,x_2|p,s|q,t)=M_{[1]}(x_1,x_2|q,t)+O(p^2)
\ee
since there is no $x_1x_2$ term by grading. This explains why, for the simplest one-box Young diagram, the Shiraishi function is an eigenfunction even at $p=0$: in all these cases, the answer is just the ordinary fundamental Macdonald polynomial.

The general eigenfunction can be realized as a finite sum
\be
\Psi_{[n]}(x_1,x_2|0,s|q,t)=\sum_{k=0}\beta_k(y_1,y_2|w|q,t)M_{[n-2k]}(x_1,x_2|q,q^{k}t)(x_1x_2)^k
\ee
with the coefficients $\beta_k(y_1,y_2|w|q,t)\sim w^k+O(w^{k+1})$. For $y_1=q^rt$, $y_2=1$ they look like
\be
\beta_k(y_1,y_2|w|q,t)=\sum_{i=0}^k \alpha_{i}^{(k)}\cdot\Big({q\over t}\Big)^{k-i}\cdot
\prod_{j=1}^{k-i}{\Theta\Big(w|{y_1\over y_2},{1\over(q^{j-1}t)^2}\Big)\over\Theta\Big(w|{y_1\over y_2},{1\over q^{2j}}\Big)}\nn\\
\alpha_i^{(k)}:=\Big(-{1\over qt}\Big)^i\cdot {q^{2i}y_1-tq^{2k}y_2\over q^iy_1-tq^{2k}y_2}\cdot
\prod_{j=1}^i q^k\cdot{y_1-tq^{2k-j}y_2\over q^{j-1}y_1-q^ky_2}\cdot{tq^{k-j}-1\over q^{j}-1}
\ee
where we have introduced the $\theta$-function of genus 2:
\be
\Theta(w|z_1,z_2):={w^{1/4}\over z_1\sqrt{z_2}}\left(\theta^{(e)}_w(z_1)\theta^{(e)}_w\Big({z_2\over w}\Big)-
\sqrt{{z_2\over z_1}}\cdot\theta^{(e)}_w(z_2)\theta^{(e)}_w\Big({z_1\over w}\Big)\right)
\ee
In the notation of \cite{BMMM}, the genus 2 $\theta$-function defined in formula (53) of \cite{BMMM} is associated with this one upon the identification
$$
s=0,\ \ w=e^{2\pi ir},\ \ z_1=e^{2\pi i(\xi_1+\xi_2)},\ \ z_2=e^{2\pi i(\xi_1-\xi_2)}
$$

Now, using formula (\ref{12}), one can lift these formulas up to the mother function:
\be
\Psi(x_1,x_2|y_1,y_2,w|q,t)=\sum_{k,n=0}\beta_k(y_1,y_2|w|q,t){x_2^{n+1+k}\over x_1^{n-k+1}}
\prod_{i=0}^n {q^{i+1}\over (q^kt)^{i+1}}{1-q^{k+i}t\over 1-q^{i+1}}{q^{3k+i}sty_2-y_1\over q^{2k+i+1}sy_2-y_1}
\ee
such that
\be
\hat{\mathfrak{O}}_0^{-1}\hat{\mathfrak{O}}_1\Big|_{p=0}\cdot x^\lambda\Psi(x_1,x_2|y_1,y_2,w|q,t)=
{y_2\over  \sqrt{t}}{\theta^{(e)}_w(y_1/(wy_2))\over\theta^{(e)}_w(y_1/y_2)}
\cdot x^\lambda\Psi(x_1,x_2|y_1,y_2,w|q,t)
\ee
which solves the eigenfunction problem for the $n=2$ dual Ruijsenaars Hamiltonian. These formulas can be straightforwardly generalized to $n>2$.

\subsection{A way to construct the eigenfunctions}

Let us explain one of the ways to construct the eigenfunctions of the Dell Hamiltonian in the two-particle case that gives the answer immediately in terms of the genus two $\theta$-functions.
It also avoids re-expansion in terms of Macdonald polynomials,
which was attempted in  (\ref{Macreexpan}).

Expanding Hamiltonians (\ref{n2hams}) in powers of $w$ and $u$,
\be
\hat{\mathfrak{O}}_0=\sum_{k,l\geq 0}
w^{k^2} u^{l(l-1)} \hat{\mathfrak{O}}_0^{(k,l)},
\ \ \ \ \ \ \  \ \ \ \ \ \ \
\hat{\mathfrak{O}}_1=\sum_{k,l\geq 1}
w^{k(k-1)} u^{l(l-1)} \hat{\mathfrak{O}}_1^{(k,l)}
\ee
one obtains very simple and instructive recurrent formulas for their action
on arbitrary symmetric functions of two $x$-variables:
\be
\!\!\!\!\!\!\!\!\!
\frac{\sqrt{x_1x_2}}{x_1-x_2}\cdot
\hat{\mathfrak{O}}_0^{(k,l)}\Big(\frac{x_1^{r+1}-x_2^{r+1}}{x_1-x_2}\cdot(x_1x_2)^m\Big)
=
\nn
\ee
\vspace{-0.5cm}
\be
= \frac{(-)^{l-1}}{1+\delta_{k,0}}\cdot {\rm sym}\Big(q^{rk}\,t^{k(2l-1)}\Big)
 \cdot \frac{x_1^{r+1}-x_2^{r+1}}{x_1-x_2}\cdot\frac{ x_1^{2l-1}-x_2^{2l-1}}{x_1-x_2}\cdot (x_1x_2)^{m+1-l} +
\nn
\ee
\vspace{-0.5cm}
\be
+ \sum_{j=1}^r
(-)^{l}\cdot {\rm asym}\!\left(q^k\right)\cdot{\rm asym}\!\left(q^{(r+1-2j)k}\,t^{k(2l-1)} \right)
 \cdot \frac{x_1^{r+1-j}-x_2^{r+1-j}}{x_1-x_2}\cdot
 \frac{ x_1^{2l-j-1}-x_2^{2l-j-1}}{x_1-x_2}\cdot (x_1x_2)^{m+j+1-l}
\nn \\
\label{bitriang}
\ee
\vspace{-0.8cm}
\be
 \!\!\!\!\!\!\!\!\!\!\!\!\!\!\!\!\!\!
\frac{\sqrt{x_1x_2}}{x_1-x_2}\cdot
\hat{\mathfrak{O}}_1^{(k,l)}\Big(\frac{x_1^{r+1}-x_2^{r+1}}{x_1-x_2}\cdot(x_1x_2)^m\Big) =
\nn
\ee
\vspace{-0.5cm}
\be
= {(-)^{l-1}}\,q^{m+\frac{r}{2}}\cdot
{\rm sym}\Big(q^{r(k-\frac{1}{2})}\,t^{(k-\frac{1}{2})(2l-1)}\Big)
 \cdot \frac{x_1^{r+1}-x_2^{r+1}}{x_1-x_2}\,
 \frac{ (x_1^{2l-1}-x_2^{2l-1})}{x_1-x_2}\cdot (x_1x_2)^{m+1-l} +
\nn
\ee
\vspace{-0.5cm}
\be
\!\!\!\!\!\!\!\!\!\!\!\!
+\sum_{j=1}^r (-)^l\cdot q^{m+\frac{r}{2}}\cdot {\rm asym}\!\left(q^{k-\frac{1}{2}}\right)\cdot
 {\rm asym}\!\left(q^{(r+1-2j)(k-\frac{1}{2})}\,t^{(k-\frac{1}{2})(2l-1)} \right)
  \cdot\frac{x_1^{r+1-j}-x_2^{r+1-j}}{x_1-x_2}\cdot
  \frac{  x_1^{2l-j-1}-x_2^{2l-j-1} }{x_1-x_2}\cdot (x_1x_2)^{m+2-l}
\nn
\ee

\bigskip

\noindent
where ${\rm sym}(x) = x+\frac{1}{x}$ and ${\rm asym}(x) = x-\frac{1}{x}$.
This is true for all integer $r,k,l\geq 0$ and for all integer $m$, not obligatory positive.
Thus, one gets a general description of bi-triangular action in the case of two $x$-variables, which is easy to sum over $k$ and $l$ and express the answer in terms
of genus two $\theta$-functions.
For $m\geq 0$, the l.h.s. can be considered
as action of $\hat{\mathfrak{O}}$-operators on an arbitrary
two-line ${\rm Schur}_{[r+m,m]}[x_1,x_2]=\frac{x_1^{r+1}-x_2^{r+1}}{x_1-x_2}\cdot(x_1x_2)^m$,
which can be straightforwardly generalized
from two to an arbitrary number of $x$-variables.
The Hamiltonians are the ratios of these triangular matrices,
but most interesting properties should be seen already at the level of (\ref{bitriang}).
The $x\leftrightarrow y$ symmetry is not yet explicit and should be revealed at further stages.

\section{Conclusion}

To summarize, in this paper, we discussed the appealing possibility that the self-dual Shiraishi series
provide eigenfunctions of the Dell systems.
We modelled the latter by the version recently advocated by P. Koroteev and Sh. Shakirov
based on the old suggestion to use higher genus theta-functions with a constant
period matrix.
Conjecturally, the dynamical period matrix, reflecting the Seiberg-Witten symplectic structure
can arise after projection from genus $n$ to $n-1$, which is a standard step in the study of
the Calogero-Ruijsenaars family systems,
but this remains to be explicitly worked out.
Anyhow, the Hamiltonians (\ref{Ham}) have a nice triangular structure, which allows a straightforward
construction of eigenfunctions through peculiar recurrent relations.
This seems indeed consistent with J. Shiraishi's anzatz, though some details
remain to be clarified.
In two Appendices below, we further comment on the relation of the entire construction to
network DIM-based models, which are widely used to build Nekrasov functions from
Dotsenko-Fateev integrals.
Once again, some effort is still needed to ``close the circle" and fully reveal the symplectic
structures and rich symmetries of the theory in the Dell case.
Hamiltonians and their eigenfunctions arise from Nekrasov functions in the $\epsilon_2=0$
(Nekrasov-Shatashvili) limit, but the Shiraishi functions can appear applicable even beyond it.
In the forthcoming paper \cite{AKMM2}, we discuss an improved version of our claim,
with an additional elliptic deformation of the Shiraishi series.

\section*{Acknowledgements}

We would like to thank M. Fukuda for discussion, in particular
for telling us the results prior to the submission to arXiv.  We are grateful to P. Koroteev, Sh. Shakirov, G. Aminov and Y.Zenkevich for enthusiasm
and useful discussions and to A. Smirnov,
who also consulted us on the modern terminology in the field.

Our work is supported in part by Grants-in-Aid for Scientific Research
(17K05275) (H.A.), (15H05738, 18K03274) (H.K.) and JSPS Bilateral Joint Projects (JSPS-RFBR collaboration)
``Elliptic algebras, vertex operators and link invariants'' from MEXT, Japan. It is also partly supported by the grant of the Foundation for the Advancement of Theoretical Physics ``BASIS" (A.Mir., A.Mor.), by  RFBR grants 19-01-00680 (A.Mir.) and 19-02-00815 (A.Mor.), by joint grants 19-51-53014-GFEN-a (A.Mir., A.Mor.), 19-51-50008-YaF-a (A.Mir.), 18-51-05015-Arm-a (A.Mir., A.Mor.), 18-51-45010-IND-a (A.Mir., A.Mor.). The work was also partly funded by RFBR and NSFB according to the research project 19-51-18006 (A.Mir., A.Mor.).  We also acknowledge the hospitality of KITP and partial support by the National Science Foundation under Grant No. NSF PHY-1748958.

\bigskip

\newpage

\section*{Appendix A. $p\to 0$ limit of the Shiraishi function and $T[U(n)]$ theory\label{p0}}

In the limit of $p \to 0$, the Shiraishi function reduces to
\beq
\lim_{p \to 0} \mathfrak{P}_n (p^{n-i} x_i ; p \vert s^{n-i} y_i ; s \vert q, \frac{q}{t})
=\sum_{\vec\lambda}
\prod_{i,j = 1}^n \frac{\mathrm{N}_{\lambda^{(i)} \lambda^{(j)}}^{(j-i)} (s^{i-j} \frac{q y_j}{t y_i} \vert q, s)}
{\mathrm{N}_{\lambda^{(i)} \lambda^{(j)}}^{(j-i)} (s^{i-j} \frac{y_j}{y_i} \vert q, s)}
\prod_{\beta=1}^n \prod_{\alpha=1}^{n-\beta} \left( \frac{t x_{\alpha+\beta}}
{q x_{\alpha+\beta-1}}\right)^{\lambda_\alpha^{(\beta)}}, \label{Mlimit}
\eeq
where we have made a change of parameter $t$ to ${q\over t}$ for the consistency with the function $P_n (x \vert y \vert q,t)$ defined
by (\ref{P}).
Note that we made the scaling $p^\delta x = (p^{n-1} x_1, \cdots, p x_{n-1}, x_n)$ and similarly for the dual variables $(y_i, s)$,
which gives an additional factor $s^{i-j}$.
Due the scaling of $x$-variables, the power $(p x_{\alpha+\beta}/ t x_{\alpha+\beta -1})^{\lambda_\alpha^{(\beta)}}$
appearing in $\mathfrak{P}_n (x ; p \vert y ; s \vert q, t)$ is scaled to
$(p^n x_{\alpha+\beta}/ t x_{\alpha+\beta -1})^{\lambda_\alpha^{(\beta)}}$, if $\alpha+\beta \equiv 1$ mod $n$ and
$(x_{\alpha+\beta}/ t x_{\alpha+\beta -1})^{\lambda_\alpha^{(\beta)}}$ otherwise.
Hence to obtain a non-vanishing result in the limit $p \to 0$ we have to impose $\lambda_\alpha^{(\beta)}=0$
for $\alpha+\beta =n+1$. Thus the length of the partition $\lambda^{(\beta)}$ is at most $n-\beta$, which is
the restriction on $\alpha$ in \eqref{Mlimit}.
As we will see later by examining the selection rule in (\ref{ellC}),
the right hand side of \eqref{Mlimit} is actually independent of the dual elliptic parameter $s$.
In \cite{S}, it is pointed out that the Shiraishi function in the limit of $p\to 0$ agrees
with the function $P_n (x \vert y \vert q,t)$ introduced in \cite{NS}
as a solution to the bispectral problem for the Ruijsenaars-Macdonald $q$-difference operators.

As we demonstrate in Appendix B, $\mathfrak{P}_n (x ; p \vert y ; s \vert q, t)$ is identified with
the Nekrasov partition function of $\mathcal{N} =2^*$ $SU(n)$ gauge theory
with the maximal monodromy defect which breaks $SU(n)$ completely to $U(1)^{n-1}$.
In the four dimensional case, the surface defect has another description by $\mathcal{N}=(2,2)$ gauged
linear sigma model coupled to the bulk theory \cite{Gaiotto:2009fs}. The coupling is achieved by gauging the flavor symmetry
and the (twisted) mass parameters are identified with the Coulomb moduli of the bulk theory.
When the bulk theory is five dimensional, we should consider $S^1$ lift of the two dimensional $\mathcal{N}=(2,2)$ theory.
In the limit of $p\to 0$, only the \lq\lq perturbative\rq\rq\ sector (the zero instanton number sector)
survives. From the viewpoint of 3d theory on the codimension two defect, this means the bulk contribution decouples.
Hence, we expect that the function $\mathfrak{P}_n (x ; p \vert y ; s \vert q, t)$  in the limit of $p \to 0$ is
identified with the vortex partition function of 3d theory. In the following we show this is indeed the case.
Namely, the function $P_n (x \vert y \vert q,t)$ agrees with
the vortex counting partition function for the holomorphic block of 3d $\mathcal{N}=4$ T[$U(n)$] theory,
where the identification of the parameters are:
\begin{center}
\begin{tabular}{|c|c|}
\hline
$P_n (x \vert y \vert q,t)$ & 3d T[$U(n)$] theory  \\ \hline
$x$ & FI parameters \\ \hline
$y$ & real mass parameters \\ \hline
$q$ & 2d $\Omega$ background \\ \hline
$t$  & axial (adjoint) mass \\ \hline
\end{tabular}
\end{center}
Note that T[$U(n)$] theory is self-mirror where the 3-dimensional mirror symmetry exchanges the FI parameters and the
chiral mass parameters. This is consistent with the fact that $P_n (x \vert y \vert q,t)$
is a solution to the bispectral problem  \cite{NS}.

T[$U(n)$] theory is a 3 dimensional $\mathcal{N}=4$ quiver gauge theory
with gauge group $ U(1) \times U(2) \times \cdots \times U(n-1)$. Originally it was introduced as a boundary theory of
4 dimensional $\mathcal{N}=4$ supersymmetric Yang-Mills theory \cite{Gaiotto:2008ak}.
The theory has bifundamental matters connecting the adjacent nodes and
$n$ hypermultiplets at the final node. Thus the flavor symmetry is $U(n)$.
In \cite{Zenkevich:2017ylb} the vortex counting partition function for the holomorphic block
of 3D $\mathcal{N}=4$ T[$U(n)$] theory is computed as follows (see also \cite{Yoshida:2014ssa}, \cite{Bullimore:2014awa}):
\beqa
Z_{\mathrm{vor}} \left(\vec{\mu}, \vec{\tau}\, \Big|\, q, \frac{q}{t}\right)
&=&
\sum_{\{ k_i^{(a)}\}} \prod_{a=1}^{n-1} \left( \frac{q \tau_{a}} {t \tau_{a+1}} \right)^{\sum_{i=1}^a k_i^{(a)}}
\prod_{i \neq j}^a \frac{\left( \frac{q \mu_i}{t \mu_j} ; q\right)_{k_i^{(a)} - k_j^{(a)}}}
{\left( \frac{\mu_i}{\mu_j} ; q\right)_{k_i^{(a)} - k_j^{(a)}}}
\prod_{i=1}^a \prod_{j=1}^{a+1} \frac{\left( \frac{t \mu_i}{\mu_j} ; q\right)_{k_i^{(a)} - k_j^{(a+1)}}}
{\left( \frac{q \mu_i}{\mu_j} ; q\right)_{k_i^{(a)} - k_j^{(a+1)}}}, \label{vortex}
\eeqa
where we have replaced $t$ in the original formula with ${q\over t}$,
which is related to the Poincar\'e duality of $P_n (x \vert y \vert q, t)$  \cite{NS}.
The parameters $\vec{\tau}$ and $\vec{\mu}$ are the (exponentiated) FI and mass parameters, respectively.
Under the 3-dimensional mirror symmetry which exchanges the Coulomb branch and the Higgs branch,
we have \cite{Zenkevich:2017ylb}:
\beq
\vec{\tau}  \leftrightarrow \vec{\mu}, \qquad t \leftrightarrow \frac{q}{t}.
\eeq
The set of non-negative integers $k_i^{(a)}~(1 \leq i \leq a)$ comes from the positions of poles in the contour
integral of screening currents and satisfies the condition:
\beq
\begin{array}{ccccccccc}
k_1^{(1)} & \geq &  k_1^{(2)} & \geq & k_1^{(3)} & \geq & \cdots  & \geq & k_1^{(n-1)}, \\
&& k_2^{(2)} &\geq & k_2^{(3)} & \geq & \cdots &\geq & k_2^{(n-1)}, \\
&&&& \ddots & \ddots & \cdots & \vdots & \vdots \\
&&&&&& k_{n-2}^{(n-2)} & \geq & k_{n-2}^{(n-1)},  \\
&&&&&&&& k_{n-1}^{(n-1)}
\end{array}
\eeq
The upper label $(a)$ of the integers $k_i^{(a)}$ stands for the color (or the $\mathbb{Z}_n$
orbifold charge) from the defect
and each row of the inequalities above corresponds to the Young diagram $\lambda^{(i)}$ with height $\ell(\lambda^{(i)}) = n-i$.
Note that the genuine holomorphic block has also classical and one-loop contributions \cite{Zenkevich:2017ylb}:
\beq
\mathcal{B}_{T[U(n)]}^{D^2 \times S^1} = Z_{\mathrm{cl}}  \left(\vec{\mu}, \vec{\tau} \,\Big| \, q, \frac{q}{t}\right)
Z_{\mathrm{1-loop}}  \left(\vec{\mu}, \vec{\tau} \,\Big| \, q, \frac{q}{t}\right)
Z_{\mathrm{vor}} \left(\vec{\mu}, \vec{\tau} \,\Big| \, q, \frac{q}{t}\right),
\eeq
where
\beq
Z_{\mathrm{1-loop}}  \left(\vec{\mu}, \vec{\tau} \,\Big| \, q, \frac{q}{t}\right) = \prod_{i<j}^n
\frac{\left( q \frac{\mu_i}{\mu_j};q \right)_\infty }{\left( t \frac{\mu_i}{\mu_j};q \right)_\infty }
\eeq
The classical part contains the theta function
\beq
 Z_{\mathrm{cl}}  \left(\vec{\mu}, \vec{\tau} \,\Big| \, q, \frac{q}{t}\right)
 \sim \prod_{i<j}^n \frac{\theta_q \left( \frac{q \mu_j}{t \mu_i} \right)}{\theta_q \left( \frac{\mu_j}{\mu_i} \right)},
\eeq
which implies some cancellations of $q$-shifted factorials between $Z_{\mathrm{1-loop}}$ and  $Z_{\mathrm{cl}}$.
The perturbative contribution $Z_{\mathrm{cl}} \cdot Z_{\mathrm{1-loop}}$ corresponds to the normalization factor
of the function (\ref{P}), which is inevitable for the bispectral duality  \cite{NS}.
It is quite remarkable that the vortex counting function $Z_{\mathrm{vor}} (\vec{\mu}, \vec{\tau} \vert q, t)$ is
obtained from the \lq\lq Higgsed\rq\rq\ network model of DIM (quantum toroidal) algebra
$U_{q,t} (\widehat{\widehat{\mathfrak{gl}_1}})$ \cite{Zenkevich:2018fzl}.
See also the computation in Appendix A of \cite{Fukuda:2019ywe}.
Hence it is natural to expect the Shiraishi function $\mathfrak{P}_n (x ; p \vert y ; s \vert q, t)$ for $p \neq 0$
can be obtained by compactifying the \lq\lq Higgsed\rq\rq\ network \cite{FOS}.
We can associate the elliptic modulus $p$ with the compactified edge,
while the appearance of the dual elliptic parameter $s$ seems rather tricky.

We can check that $P_n (x \vert s \vert q,t)$ agrees with
$Z_{\mathrm{vor}}( \vec{\mu}, \vec{\tau} \vert q, \frac{q}{t} )$ with the relation
\beq
m_{ij} = \lambda_{j-i}^{(i)} - \lambda_{j-i+1}^{(i)} = k_i^{(j-1)} - k_i^{(j)}
\eeq
Substituting this relation, we obtain
\beqa
&& C_n( k_i^{(a)} ; y \vert q,t)  \CR
&=& \prod_{a=1}^{n-1} \prod_{1 \leq i < j \leq a+1} \frac{(q^{k_i^{(a+1)} - k_j^{(a+1)}} \frac{t y_j}{y_i} ; q) _{k_i^{(a)} - k_i^{(a+1)}}}
{(q^{k_i^{(a+1) - k_j^{(a+1)}}} \frac{q y_j}{y_i} ; q)_{k_i^{(a)} - k_i^{(a+1)}}}
\prod_{1 \leq i \leq j \leq a}
\frac{(q^{-k_j^{(a)} - k_i^{(a+1)}} \frac{q y_j}{t y_i} ; q) _{k_i^{(a)} - k_i^{(a+1)}}}
{(q^{-k_j^{(a) - k_i^{(a+1)}}} \frac{y_j}{y_i} ; q)_{k_i^{(a)} - k_i^{(a+1)}}},
\eeqa
where we have set $a=k-1$.
We see that the factors in $C_n$ with $1 \leq i=j \leq a$ are
\beq
\frac{(q^{-\theta_{ik}} \frac{q}{t} ; q)_{\theta_{ik}}}{(q^{-\theta_{ik}}; q)_{\theta_{ik}}}
= \left( \frac{q}{t} \right)^{k_i^{(a)} - k_i^{(a+1)} }
\frac{ (t;q)_{k_i^{(a)} - k_i^{(a+1)}}}{ (q;q)_{k_i^{(a)} - k_i^{(a+1)}}}
\eeq
We find up to the power of ${q\over t}$, the corresponding factors in $Z_{\mathrm{vor}}$ are exactly the same as above.
When $i < j$, we use the formula
\beq
(q^m u ; q)_n = \frac{(u;q)_{m+n}} {(u ; q)_m}, \qquad m, n \in \mathbb{Z}
\eeq
which is valid also for negative integers. Then we find the following factors:
\beq
\frac{ \left( \frac{t y_j}{y_i} ; q \right)_{k_i^{(a)} - k_j^{(a+1)}}}{ \left( \frac{q y_j}{y_i} ; q\right)_{k_i^{(a)} - k_j^{(a+1)}}}
\frac{ \left( \frac{q y_j}{y_i} ; q\right)_{k_i^{(a+1)} - k_j^{(a+1)}}}{ \left( \frac{t y_j}{y_i} ; q \right)_{k_i^{(a+1)} - k_j^{(a+1)}}},
\qquad 1 \leq i < j \leq a+1, \label{factor1}
\eeq
and
\beq
\frac{ \left( \frac{q y_j}{t y_i} ; q \right)_{k_i^{(a)} - k_j^{(a)}}}{ \left( \frac{y_j}{y_i} ; q\right)_{k_i^{(a)} - k_j^{(a)}}}
\frac{ \left( \frac{y_j}{y_i} ; q\right)_{k_i^{(a+1)} - k_j^{(a)}}}{ \left( \frac{q y_j}{t y_i} ; q \right)_{k_i^{(a+1)} - k_j^{(a)}}}, \label{factor2}
\qquad 1 \leq i < j \leq a
\eeq
In each case we see the first factor agrees with the factors in $Z_{\mathrm{vor}}$ with $i<j$ by substituting $y_i = \mu_i^{-1}$,
where we have taken the condition $k_i^{(n)} =0$ into account.
To obtain the missing factors with $j<i$, we exchange $i$ and $j$ in the second factors of \eqref{factor1} and \eqref{factor2}.
If we employ the formula:
\beq
(u ; q)_n = (-u)^{n} q^{\frac{1}{2}n(n-1)} (q u^{-1} ;q)_{-n}^{-1}, \qquad n \in \mathbb{Z}_{\geq 0}.m  \label{flip}
\eeq
the second factor gives
\beq
\left(\frac{q}{t}\right)^{k_j^{(a+1)} - k_i^{(a+1)}}
\frac{ \left( \frac{q y_j}{t y_i} ; q \right)_{k_i^{(a+1)} - k_j^{(a+1)}}} { \left( \frac{y_j}{y_i} ; q\right)_{k_i^{(a+1)} - k_j^{(a+1)}}}
\qquad  1 \leq j < i \leq a+1
\eeq
and
\beq
\left(\frac{t}{q}\right)^{k_j^{(a+1)} - k_i^{(a)}}
\frac{ \left( \frac{t y_j}{ y_i} ; q \right)_{k_i^{(a)} - k_j^{(a+1)}}}{ \left( \frac{q y_j}{y_i} ; q\right)_{k_i^{(a)} - k_j^{(a+1)}}},
\qquad 1 \leq j < i \leq a
\eeq
respectively.
Taking the condition $k_i^{(n)} =0$ into account again, we can see these factors indeed
give the missing factors for $j<i$ in  $Z_{\mathrm{vor}}$, up to the power of ${q\over t}$.
Finally for completeness let us count the total power of ${q\over t}$ that arose during the above computations:
\beqa
&&\sum_{a=1}^{n-1} \sum_{i=1}^a (k_i^{(a)} - k_i^{(a+1)}) + \sum_{a=1}^{n-1} \sum_{1 \leq j < i \leq a+1} (k_j^{(a+1)} - k_i^{(a+1)})
+ \sum_{a=1}^{n-1} \sum_{1 \leq j < i \leq a} (k_i^{(a)} - k_j^{(a+1)}) \CR
&=& \sum_{a=1}^{n-1} k_{a}^{(a)} + \sum_{a=1}^{n-1} \sum_{j=1}^a k_j^{(a+1)}
-  \sum_{a=1}^{n-2} \sum_{1 \leq j < i \leq a+1} k_i^{(a+1)} + \sum_{a=2}^{n-1} \sum_{1 \leq j < i \leq a} k_i^{(a)} \CR
&=& \sum_{a=1}^{n-1} \sum_{i=1}^{a}  k_i^{(a)} \label{power}
\eeqa
where we have used $k_i^{(n)}=0$. Hence the power is exactly the same as that of \eqref{Mlimit}.

Armed with the agreement of the Noumi-Shiraishi representation of the Macdonald function $P_n (x \vert y \vert q,t)$ and
the vortex partition function of $T[U(n)]$ theory, we can show that they also agree with \eqref{Mlimit}.
We first have to examine the selection rule in the Nekrasov factor (\ref{ellC}).
Since $1 \leq \alpha \leq \beta \leq n-1$ in the limit $p \to 0$,
when $k\geq 0$, there is no solution to the selection rule in the second factor of (\ref{ellC})
and there is a unique solution $\beta = \alpha + k$ to the selection rule in the first factor.
On the other hand when $k < 0$, it is the second factor that has a unique solution $\beta = \alpha - k-1$
and the first factor has no solution to the selection rule.
Hence, substituting the relation $\lambda_\ell^{(i)} = k_i^{(\ell+i-1)}$ we obtain the following three contributions:
\begin{enumerate}
\item $i=j$
\beq
\prod_{i=1}^{n} \frac{\mathrm{N}_{\lambda^{(i)}, \lambda^{(i)}}^{(0)} \left( \frac{q}{t}  \vert q, s \right)}
{\mathrm{N}_{\lambda^{(i)}, \lambda^{(i)}}^{(0)} \left( 1 \vert q, s \right)}
= \prod_{i=1}^{n-1} \prod_{\alpha=1}^{n-i} \frac{(t; q)_{k_i^{(i+\alpha-1)} - k_i^{(i+\alpha)}}}{(q; q)_{k_i^{(i+\alpha-1)} - k_i^{(i+\alpha)}}}
\left( \frac{q}{t} \right)^{k_i^{(i+\alpha-1)} - k_i^{(i+\alpha)}}
\eeq

\item $i<j$
\beqa
&&\prod_{1 \leq i < j \leq n}  \frac{\mathrm{N}_{\lambda^{(i)}, \lambda^{(j)}}^{(j-i)}
\left( s^{i-j} \frac{q y_j}{t y_i}  \vert q, s \right)}
{\mathrm{N}_{\lambda^{(i)}, \lambda^{(j)}}^{(j-i)} \left( s^{i-j} \frac{y_j}{y_i} \vert q, s\right)} \CR
&=&  \prod_{1 \leq i < j \leq n}  \prod_{\alpha=1}^{n-j}
\frac{( \frac{q y_j}{t y_i}  ; q)_{k_i^{(\alpha + j -1)} - k_j^{(\alpha+ j-1)}}}
{( \frac{q y_j}{t y_i}  ; q)_{k_i^{(\alpha + j )} - k_j^{(\alpha+ j-1)}}}
\frac{( \frac{y_j}{y_i}  ; q)_{k_i^{(\alpha + j )} - k_j^{(\alpha+ j-1)}}}
{( \frac{y_j}{y_i}  ; q)_{k_i^{(\alpha + j -1)} - k_j^{(\alpha+ j-1)}}}.
\eeqa

\item $i>j$
\beqa
&&\prod_{1 \leq j < i \leq n} \frac{\mathrm{N}_{\lambda^{(i)}, \lambda^{(j)}}^{(j-i)}
\left( s^{i-j} \frac{q s_j}{t s_i}  \vert q, s \right)}
{\mathrm{N}_{\lambda^{(i)}, \lambda^{(j)}}^{(j-i)} \left( s^{i-j} \frac{s_j}{s_i} \vert q, s \right)} \CR
&=&  \prod_{1 \leq j < i \leq n}  \prod_{\alpha=1}^{n-i+1}
\frac{( \frac{q y_j}{t y_i} ; q)_{k_i^{(\alpha + i -1)} - k_j^{(\alpha+ i -1)}}}
{( \frac{q y_j}{t y_i}; q)_{k_i^{(\alpha + i -1)} - k_j^{(\alpha+ i-2)}}}
\frac{( \frac{y_j}{y_i}; q)_{k_i^{(\alpha + i -1)} - k_j^{(\alpha+ i-2)}}}
{( \frac{y_j}{y_i} ; q)_{k_i^{(\alpha + i -1)} - k_j^{(\alpha+ i-1)}}}
\eeqa

\end{enumerate}
In the first case,  setting $a = i +\alpha-1$, we have $1 \leq a \leq n-1$ and $1 \leq i \leq a$. Hence
the contribution becomes
\beq
\prod_{a=1}^{n-1} \prod_{i=1}^{a} \frac{(t; q)_{k_i^{(a)} - k_i^{(a+1)}}}{(q; q)_{k_i^{(a)} - k_i^{(a+1)}}}
\left( \frac{q}{t} \right)^{k_i^{(a)} - k_i^{(a+1)}}.
\eeq
In the second case, setting $a = j + \alpha -1$ implies $2 \leq a \leq n-1$ and $1 \leq i <  j \leq a$.
Hence we obtain
\beq
\prod_{a=2}^{n-1} \prod_{1 \leq i < j \leq a}
\frac{( \frac{q y_j}{t y_i}  ; q)_{k_i^{(a)} - k_j^{(a)}}}
{( q \frac{y_i}{y_j}  ; q)_{k_j^{(a)} - k_i^{(a+1)}}}
\frac{( t \frac{y_i}{y_j}  ; q)_{k_j^{(a)} - k_i^{(a+1)}}}
{( \frac{y_j}{y_i}  ; q)_{k_i^{(a)} - k_j^{(a)}}}
\left( \frac{q}{t} \right)^{k_j^{(a)} -k_i^{(a+1)}},
\eeq
where we have used \eqref{flip}.
Finally in the last case, setting $a = i + \alpha -2$ implies $1 \leq a \leq n-1$ and $1 \leq j < i \leq a+1$
\beq
\prod_{a=1}^{n-1} \prod_{1 \leq j < i \leq a+1}
\frac{( \frac{q y_j}{t y_i} ; q)_{k_i^{(a+1)} - k_j^{(a+1)}}}
{( q \frac{y_i}{y_j}; q)_{k_j^{(a )} - k_i^{(a+1)}}}
\frac{( t \frac{y_i}{y_j}; q)_{ k_j^{(a )} - k_i^{(a+1)}}}
{( \frac{y_j}{y_i} ; q)_{k_i^{(a+1)} - k_j^{(a+1)}}}
\left( \frac{q}{t} \right)^{k_j^{(a)} - k_i^{(a+1)}},
\eeq
where we have used \eqref{flip} again. Then by the same change of variables $y_i = 1/\mu_i$ as before
we can find an agreement with \eqref{vortex} up to the power of ${q\over t}$. Note that we have to exchange $i$ and $j$
for the factors with  $k_j^{(a)} - k_i^{(a+1)}$.
Finally one can check the total power of ${q\over t}$ is correct by a similar counting to \eqref{power}.

%

\subsection*{Appendix B. Shiraishi function and maximal monodromy defect}

Let us note that the power of the series expansion (\ref{ellP}) can also be rewritten as
\beq
\prod_{\beta=1}^n \prod_{\alpha \geq 1}
\left( \frac{p x_{\alpha + \beta}}{t x_{\alpha + \beta -1}} \right)^{\lam_\alpha^{(\beta)}}
= \left( \frac{p}{t} \right)^{|\vec{\lambda}|} \prod_{i=1}^n x_i^{m_i},
\eeq
where $m_i = d_{i} - d_{i +1}$ with
\beq
|\vec{\lambda}| := \sum_{\beta=1}^n |\lambda^{(\beta)}|, \qquad
d_i (\vec{\lambda}):= \sum_{\alpha=1}^\infty
\sum_{\alpha + \beta \equiv i  \atop (\mathrm{mod}~n)} \lambda_\alpha^{(\beta)}
\eeq
The integer $m_i$ with $\displaystyle{\sum_{i=1}^n} m_i =0$ corresponds
to the magnetic flux associated with the monodromy defect which breaks
$SU(n)$ to $U(1)^{n-1}$ \cite{Kanno:2011fw}.
In \cite{S} it was pointed out $\mathfrak{P}_n (x_i ; p \vert y_i ; s \vert q,t)$
is identified with the equivariant Euler characteristic of the affine Laumon space \cite{FFNR},
while in \cite{KS} it was argued that the eigenfunction of elliptic integrable system is
related to the instanton partition function with monodromy defect, which in turn is obtained from
the ordinary instanton partition function by introducing appropriate $\mathbb{Z}_n$-orbifold action
on the equivariant parameters \cite{Braverman:2010ef}, \cite{FR}, \cite{Kanno:2011fw}, \cite{Nawata:2014nca},
\cite{Bullimore:2014awa}, \cite{Nekrasov:2017rqy}.

In the following we summarize how
the orbifold action correctly reproduces  the equivariant Euler characteristic of the affine Laumon space
derived in \cite{FFNR}.
In fact, at the level of the equivariant character to be discussed later, the selection rules $ j - i \equiv k~(\mathrm{mod}~n)$
and $\beta  - \alpha \equiv -k-1~(\mathrm{mod}~n)$ mean
taking the terms with the charge $k/n$, if we assign the fractional charge $1/n$ for the orbifold action of $\mathbb{Z}_n$
to the parameter $s$. Hence if we define the charge of the Coulomb moduli parameter $y_i$  to be $-i/n$, then
the function $\mathfrak{P}_n (x_i ; p \vert y_i ; s \vert q,t)$ corresponds to the neutral (integral) charge sector of
the equivariant character.

Let us first confirm the Nekrasov factor (\ref{ellC}) {\it without the selection rules}
\beq
\mathrm{N}_{\lam, \mu}(u \vert q, 1/s) =
\prod_{j \geq i \geq 1} (uq^{-\mu_i + \lam_{j+1}} s^{i - j} ;q)_{\lam_j - \lam_{j +1}} \cdot
\prod_{\beta \geq \alpha \geq 1} (uq^{\lam_\alpha - \m_\beta} s^{\beta - \alpha +1} ;q)_{\m_\beta - \m_{\beta+1}},
\eeq
agrees with the standard one (see e.g. \cite{Awata:2008ed}). Using (\ref{Poch}), we obtain
\beqa
\mathrm{N}_{\lam, \mu}(u \vert q, 1/s) &=&
\prod_{j \geq i \geq 1} \frac{(uq^{\lam_{j+1} -\mu_i} s^{i - j} ;q)_\infty}
{(uq^{\lam_{j} -\mu_i} s^{i - j} ;q)_\infty}
\ \cdot
\prod_{i \geq j \geq 1} \frac{(uq^{\lam_j - \m_i} s^{i - j +1} ;q)_\infty}
{(uq^{\lam_j - \m_{i+1}} s^{i - j +1} ;q)_\infty} \ = \CR
&=&
\prod_{i, j =1}^\infty
\frac{(uq^{\lam_{j} -\mu_i} s^{i - j +1} ;q)_\infty}
{(uq^{\lam_{j} -\mu_i} s^{i - j} ;q)_\infty}
\frac {(u s^{i - j} ;q)_\infty} {(u  s^{i - j +1} ;q)_\infty} \ =\CR
&=& \exp \left( \sum_{n=1}^\infty \frac{u^n}{n} \frac{1-s^n}{1 - q^n} \left[ p_n(q^{\lam_j} s^{-j} ) p_n(q^{- \m_i} s^{i})
- p_n(s^{-j}) p_n( s^{i}) \right] \right) \label{plethystic}
\eeqa
where $p_n(\bullet)$ is the power sum function. Thus we can see $\mathrm{N}_{\lam, \mu}(u \vert q, 1/s)$ agrees with
the standard Nekrasov factor in terms of the power sum functions.
We note that the equivariant parameters for the $\Omega$ background is $q_1= e^{\epsilon} = q$ and
$q_2 = e^{\epsilon_2} = s^{-1}$ and $t$ does not correspond to the $\Omega$ background: physical meaning of the parameter $t$ is the mass of the adjoint matter $t=e^{-m}$.
On the other hand when we drive the Macdonald function from $\mathfrak{P}_n (x_i ; p \vert y_i ; s \vert q,t)$
the deformation parameters are actually $(q,t)$. Thus there is \lq\lq a mismatch\rq\rq\ between the deformation
parameters of the Macdonald function and the $\Omega$ background.

Geometrically the Nekrasov factor $\mathrm{N}_{\lam, \mu}(u \vert q, s)$
is derived from the equivariant character of the tangent space of
the instanton moduli space $T_{\vec\lam} \mathcal{M}$ at the isolated fixed points of the torus action,
which are labelled by $n$-tuples of Young diagrams $\vec\lam = \{ \lam^\alpha \}$.
According to \cite{Nakajima}, \cite{Bruzzo:2002xf}, the relevant equivariant character is given by:
\beq
\chi (u_\alpha ; q_i)  = N^{*} K + q_1 q_2 K^{*} N - (1-q_1)(1-q_2) K^{*} K, \label{eqcha}
\eeq
where\footnote{
Compared with the standard formula,
we have exchanged $q_1$ and $q_2$, or take the transpose of the Young diagram.}
\beq
N := \sum_{\alpha=1}^n u_\alpha, \qquad K :=  \sum_{\alpha=1}^n u_\alpha \cdot
\left( \sum_{(i,j) \in \lam^\alpha} q_1^{1-j} q_2^{1-i} \right)
\eeq
$N^{*}$ and $K^{*}$ denote dual characters.
$u_\alpha$ are coordinates of the Cartan torus of the gauge group $U(n)$
and $q_i$ are equivariant parameters of the torus action on $\mathbb{C}^2$.

The equivariant character of the tangent space at the fixed points of the affine Laumon space
is given by $\mathbb{Z}_n$ invariant part of the character by introducing the orbifold action on the
equivariant parameters $(u_\alpha, q_i)$.
Thus the denominator of the Shiraishi function $\mathfrak{P}_n (x_i ; p \vert y_i ; s \vert q,t)$ is
related to the equivariant character of the affine Laumon space \cite{S},
which is identified with the instanton moduli space with the maximal monodromy defect corresponding to the partition $N = (1^n)$,
which breaks $U(n)$ completely to $U(1)^n$.
The CFT side of the AGT relation in this case is supposed to be the conformal block of
the affine algebra $\widehat{\mathfrak{sl}}_n$ \cite{Alday:2010vg},\cite{Kozcaz:2010yp}.

Let us show $\mathbb{Z}_n$ invariant part of the equivariant character \eqref{eqcha} actually gives
the character formula \cite{FFNR}
\beqa
&& \mathrm{Ch}_{(\vec\lam, \vec\mu)}
 [(\vec{a}, \vec{b}) ;  q_1, q_2 ] \, = \CR
&=&
( 1- q_1) \sum_{k=1}^n \sum_{1 \leq \ell} \sum_{1 \leq \tilde\ell}
e^{a_{k-\ell +1} - b_{k -\tilde\ell}} q_2^{\left( \floor{\frac{\tilde\ell -k}{n}} - \floor{\frac{\ell -k -1}{n}}\right)}
\sum_{i=1}^{\m^{(k - \tilde\ell)}_{\tilde\ell}} q_1^{i-1} \sum_{j=1}^{\lam^{(k - \ell +1)}_\ell} q_1^{1-j} \ + \CR
&& ~~ +\, q_1 \sum_{k=1}^n \sum_{ 1 \leq \tilde \ell } e^{a_k - b_{k -\tilde\ell}}
q_2^{\left(\floor{\frac{\tilde\ell-k}{n}} - \floor{-\frac{k}{n}}\right)}
\sum_{i=1}^{\m^{k-\tilde\ell}_{\tilde\ell}} q_1^{i-1} \ -\CR
&& -\, ( 1- q_1)  \sum_{k=1}^n  \sum_{ 1 \leq \ell } \sum_{1 \leq \tilde\ell}
e^{a_{k-\ell +1} - b_{k -\tilde\ell +1}}
q_2^{\left( \floor{\frac{\tilde\ell -k -1}{n}} - \floor{\frac{\ell -k -1}{n}}\right)}
\sum_{i=1}^{\m^{(k - \tilde\ell +1)}_{\tilde\ell}} q_1^{i-1} \sum_{j=1}^{\lam^{(k - \ell +1)}_\ell} q_1^{1-j}  \ + \CR
&& +\, \sum_{k=1}^n  \sum_{1 \leq \ell} e^{a_{k-\ell +1} - b_{k}}
q_2^{\left( \floor{\frac{-k}{n}} - \floor{\frac{\ell -k -1}{n}}\right)} \sum_{j=1}^{\lam^{(k - \ell +1)}_\ell} q_1^{1-j}
\eeqa
where we have made a change of variables $\ell \to k - \ell +1$ and
$\tilde\ell \to k - \tilde\ell$ (but $\tilde\ell \to k - \tilde\ell +1$ only for the third term)
in the original formula  (Prop. 4.15 in \cite{FFNR}).
%
Multiplied with $e^{-m}$, \eqref{BFNR} gives the character for a bifundamental matter with mass $m$.
To get an adjoint matter we specialize $a=b$ and $\lam=\m$.
The character for the vector multiplet is obtained from that of adjoint matter by setting $m=0$ and reversing the overall sign.


Replacing $\ell \to  mn + \ell$ and $ \tilde\ell \to  \tilde m n + \tilde\ell$
with $0 \leq m,  \tilde m$ and $1 \leq \ell, \tilde\ell \leq n$, we can rewrite the character as follows:
\beqa
  \mathrm{Ch}_{(\vec\lam, \vec\mu)} [(\vec{a}, \vec{b}) ;  q_1, q_2 ]
&=& (1- q_1) \sum_{k=1}^n  V_{k-1}^{*} (\vec{b}, \vec\m) V_k (\vec{a}, \vec\lam)
+ q_1  \sum_{k=1}^n V_{k-1}^{*}  (\vec{b}, \vec\m) W_k (\vec{a}) - \CR
&&~~ - \sum_{k=1}^n   (1-q_1) V_k^{*} ( \vec{b}, \vec\m) V_k (\vec{a}, \vec\lam)
+  \sum_{k=1}^n W_k^{*} (\vec{b}) V_k (\vec{a}, \vec\lam),
\label{Fformula}
\eeqa
where
$
W_k (\vec{a}) := e^{a_k} q_2
$
and
\beqa
V_k  (\vec{a}, \vec\lam) &:=& \sum_{0 \leq m}  \sum_{\ell =1}^n e^{a_{k-\ell +1}}
q_2^{-m - \floor{\frac{\ell -k -1}{n}}} \sum_{j=1}^{\lam^{(k - \ell +1)}_{mn+ \ell}} q_1^{1-j} =
 \CR
&=&  \sum_{0 \leq m} \left(\sum_{\ell =1}^k e^{a_{k-\ell +1}}  \sum_{(i,mn + \ell) \in \lam^{(k-\ell+1)}} q_1^{1-i} q_2^{-m+1}
+ \sum_{\ell = k+1}^n e^{a_{k-\ell +1}}  \sum_{(i, mn +\ell) \in \lam^{(k-\ell+1)}} q_1^{1-i} q_2^{-m}  \right) \ \ \ \  \ \ \ \ \
\label{BFNR}
\eeqa
To eliminate the floor function in the formula \eqref{BFNR} we use that fact that when $1 \leq k, \ell,  \tilde\ell  \leq n$
the arguments $X$ in the floor function appearing the formula satisfies $-1 \leq X < 1$.
Therefore we have either $\floor{X} =0$ for $0 \leq X < 1$ or $\floor{X} = -1$ for $-1 \leq X < 0$, respectively.
Then we can see that \eqref{Fformula} is nothing but the $\mathbb{Z}_n$ invariant (the charge zero) part of\footnote{
Compare it with \eqref{eqcha}.}
\beq
\chi_{(\vec\lam, \vec\m)} (a_\alpha, b_\alpha; q_i)
= - (1- q_1)(1 - q_2^{\frac{1}{n}}) V_{\vec{b}, \vec{\m}}^{*}
\otimes V_{\vec{a}, \vec{\lam}} + W_{\vec{b}}^{*} \otimes V_{\vec{a}, \vec{\lam}}
+ q_1 q_2^{\frac{1}{n}} V_{\vec{b}, \vec{\m}}^{*} \otimes W_{\vec{a}},
\eeq
namely
\beqa
\chi^{\mathbb{Z}_n}_{(\vec\lam, \vec\m)} (a_\alpha, b_\alpha; q_i)
&=&
- (1- q_1) \sum_{k=1}^n (V_{\vec{b}, \vec{\m}}^{*})_k \otimes (V_{\vec{a}, \vec{\lam}})_k
+ (1- q_1) \sum_{k=1}^n (V_{\vec{b}, \vec{\m}}^{*})_{k-1} \otimes (V_{\vec{a}, \vec{\lam}})_k \CR
&&~~ + \sum_{k=1}^n (W_{\vec{b}}^{*})_k \otimes (V_{\vec{a}, \vec{\lam}})_k
+  q_1 \sum_{k=1}^n (V_{\vec{b}, \vec{\m}}^{*})_{k-1} \otimes (W_{\vec{a}})_k , \label{ZNinv}
\eeqa
where $W_n \equiv W_0$ and $V_n \equiv V_0$ with
\beq
(W_{\vec{a}})_k = e^{a_k} q_2^{1- \frac{k}{n}}, \label{W}
\eeq
and
\be
(V_{\vec{a}, \vec{\lam}})_k
&=& \sum_{\ell=0}^{k-1} e^{a_{k-\ell}} q_2^{1 - \frac{k-\ell}{n}}
\left( \sum_{(i, n(j-1) + \ell +1) \in \lam^{(k-\ell)}}
q_1^{1-i} q_2^{- (j-1) - \frac{\ell}{n}} \right) + \CR
&& +~~ \sum_{\ell= k}^{n-1} e^{a_{k-\ell+n}} q_2^{ - \frac{k-\ell}{n}}
\left( \sum_{(i, n(j-1) + \ell +1) \in \lam^{(k-\ell+n)}}
q_1^{1-i} q_2^{- (j-1)- \frac{\ell}{n}} \right)  =
\label{V}
\ee
\vspace{-0.5cm}
\be
=\ \ \sum_{\ell=1}^{k} e^{a_{k-\ell+1}} q_2^{1- \frac{k}{n}}
\left( \sum_{(i, m n + \ell) \in \lam^{(k-\ell+1)}}
q_1^{1-i} q_2^{- m} \right)
~~ +~~ \sum_{\ell= k+1}^{n} e^{a_{k-\ell+n+1}} q_2^{- \frac{k}{n}}
\left( \sum_{(i, m n + \ell) \in \lam^{(k-\ell+n+1)}}
q_1^{1-i} q_2^{- m} \right)
\nn
\ee
Note that the $\mathbb{Z}_n$ fractional charge of $W_k$ and $V_k$ defined by \eqref{W} and \eqref{V} is $(1 - k/n)$
and we have rescaled them by multiplying $q_2^{k/n}$ so that they have unit charge.

It is known that there are two ways of computing the instanton partition function with a monodromy (surface) defect
\cite{BE,Jeong:2017pai,Jeong:2018qpc}.
One is the orbifold construction described above and the other is the degenerate gauge vertex construction
in the quiver gauge theory, where we tune the Coulomb moduli and mass parameters\footnote{In the AGT dictionary,
this corresponds to the insertion of a fully degenerate primary field.} \cite{AY,MT,MMM,Awata:2010bz,Wyllard:2010vi}. It was argued that the two constructions are related
by the brane transition in $M$ theory so that they are dual in IR \cite{Frenkel:2015rda}.
The fact that the Shiraishi function $\mathfrak{P}_n (x_i ; p \vert y_i ; s \vert q,t)$
agrees with the holomorphic block of $T[U(n)]$ theory in $p \to 0$ limit is in accord with the second description
in terms of the quiver gauge theory. Note that the network diagram for $T[U(n)]$ theory is what is called
Higgsed network in \cite{Zenkevich:2018fzl}.

\end{document}